\documentclass[prl, twocolumn, showpacs, showkeys, preprintnumbers,
  amsmath,amssymb, aps, floatfix]{revtex4-1}

\usepackage{graphicx}
\usepackage{float}
\usepackage{color}
\usepackage{nicefrac}
\usepackage{wrapfig}
\usepackage{epstopdf}

\begin{document}

\title{Large non-collinearity and spin-reorientation in the novel Mn$_2$RhSn Heusler magnet}

\author{O.~Meshcheriakova$^{1,2}$, S.~Chadov$^2$, A.~K.~Nayak$^2$, U.~K.~R\"o\ss ler$^{3}$,
  J.~K\"ubler$^4$, G.~Andr\'{e}$^5$,
  A.~A.~Tsirlin$^2$, J.~Kiss$^2$, S.~Hausdorf$^2$, A.~Kalache$^2$, W.~Schnelle$^2$, M.~Nicklas$^2$, C.~Felser$^2$}
\affiliation{$^1$Graduate~School~of~Excellence~''Materials~Science~in~Mainz'' Johannes Gutenberg - Universtit\"{a}t,  55099 Mainz, Germany}
\affiliation{$^2$Max-Planck-Institut f\"ur Chemische Physik fester Stoffe,  N\"othnitzer Str.~40, 01187 Dresden, Germany}
\affiliation{$^3$Leibniz-Institut f\"{u}r Festk\"orper- und  Werkstoffforschung IFW, Helmholtz Str.~20, 01171 Dresden, Germany}
\affiliation{$^4$Institut f\"{u}r Festk\"{o}rperphysik, Technische  Universit\"{a}t Darmstadt, Germany}
\affiliation{$^5$Laboratoire~L\'eon~Brillouin,~CEA-CNRS~Saclay,~Gif-sur-Yvette~Cedex,~France}

\begin{abstract}
 Non-collinear magnets provide  essential  ingredients for the next generation memory technology.
It is a new prospect for the Heusler materials, already well-known  due to the diverse range of other fundamental characteristics.
Here we  present combined experimental/theoretical study of novel non-collinear tetragonal  Mn$_2$RhSn Heusler
material exhibiting  unusually strong canting of its magnetic sublattices.  It  undergoes a  spin-reorientation transition,
 induced by a temperature change and suppressed by an external  magnetic   field.  Due to  the   presence of
Dzya\-lo\-shin\-skii-Mo\-ri\-ya exchange and magnetic anisotropy,  Mn$_2$RhSn is suggested to be a promising candidate for realizing the
 skyrmion state in the Heusler family.
\end{abstract}
\maketitle

\indent
The art  of controlling magnetic degrees  of freedom has led  to a broad
range  of applications  that make  up  the rapidly  developing field  of
spintronics.  Up to  now,  most  of the  exploited  compounds have  been
so-called collinear magnets, i.e.   materials in which the magnetization
is formed by local magnetic  moments aligned parallel or antiparallel to
one  another.    Yet,  the  possibility  of   influencing  their  mutual
orientation   opens  new   horizons  for   the  field   of  spintronics.
Non-collinear   magnets  can  be   widely  applied   in  current-induced
spin-dynamics~\cite{BGB09},    magnetic   tunnel   junctions~\cite{YKM11},
molecular   spintronics~\cite{SC10},   spin-torque   transfer  by   small
switching     currents~\cite{CJT+08}  and     anomalous     exchange
bias~\cite{TDL+13}.  Impressive  improvement   of  the  critical  current
density by five orders of  magnitude~\cite{JMP+10,YKZ+12,SRB+12} is
offered by non-collinear magnets driven into the skyrmion phase~\cite{BH94,MBJ+09,UKR06,JMP+10,SYIT12,YKZ+12,YMY12,SRB+12,MKS+13}.
While  such  exotic magnetic  arrangements  are  sensitive  to
external conditions (magnetic  field and temperature), an expansion
of the related material base is important for their stabilization.

Flexible tuning of the magnetic properties can ultimately be realized in
multicomponent  systems  of  several magnetic  sublattices  with
competing types  of interactions such  as magneto-crystalline anisotropy,
dipole-dipole and Dzyaloshinsky-Moriya (DM)
interactions~\cite{Dzy58,Mor60}. Heusler  compounds, of  which there
are over 1000  members, provide a rich variety  of parameters for almost
any      material     engineering      task      (e.g. half-metallic
ferromagnetism~\cite{KWS83,GMEB83},  shape memory~\cite{KII+06},
exchange bias~\cite{NNC+13},    topological  insulators~\cite{CQK+10},
spin-gapless   semiconductivity~\cite{OFFK13},   spin-resolved   electron
localization~\cite{CKF13} and superconductivity~\cite{KWG+12}).
Furthermore, the majority  of Mn$_2$YZ (Y -- transition
metal, Z -- main-group atom) systems are non-centrosymmetric;
this together with the magneto-crystalline anisotropy induced by intrinsic tetragonal distortion
makes such systems attractive for skyrmion research.

First, we will discuss here the unusual ground-state magnetic canting
observed  in Mn$_2$RhSn together with the subsequent temperature-induced
spin-reorientation into the collinear ferrimagnetic mode. Further, we will
give a detailed micromagnetic analysis which suggests this collinear regime to
provide perfect conditions for the skyrmion formation, in agreement with
the earlier theoretical studies~\cite{BY89,BH94,BRWM02}.

\begin{figure}[htb]
  \centering
  \includegraphics[width=0.9\linewidth,clip]{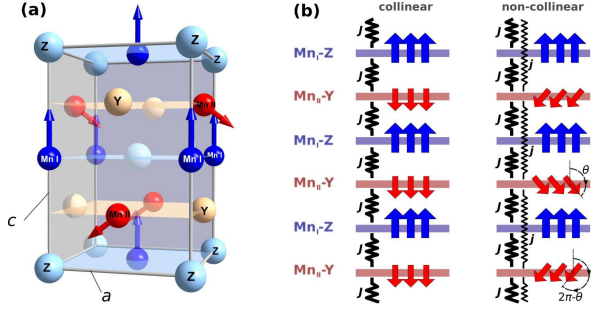}
  \caption{(a)~Crystal and magnetic structures of Mn$_2$YZ Heusler
    compounds. Due to the magneto-crystalline anisotropy induced by the
    tetragonal distortion, the Mn$_{\,\rm I}$ magnetic moments are oriented
    along the $c$ axis; the moments on Mn$_{\rm II}$ are canted in an
    alternating manner with respect to the  $c$ axis.  (b)~Schematic picture  of  the leading  magnetic
    exchange  interactions between different  atomic layers  in Mn$_2$YZ
    (atomic planes containing  Z and Y elements  are  shown in blue
    and  red, respectively).  The  arrows show  the orientation  of the spin
    moments  on Mn  and the  springs show the exchange interactions between
    different planes.  Considering only the  nearest antiparallel
    interactions $J$ (between  Mn$_{\,\rm  I}$-Z and  Mn$_{\rm  II}$-Y
    planes) leaves  the magnetic structure  collinear;
     introducing  the next-nearest  antiparallel  coupling $j$
    (between Mn$_{\rm  II}$-Y planes) leads  to the alternating canting  of Mn$_{\rm
      II}$ moments by $\theta$ and ${2\pi-\theta}$.
\label{fig:jxc}}
\end{figure}

 In  a non-relativistic case,  the magnetic non-collinearity is a
 result of the  competition between antiparallel and parallel exchange interactions (or between several types
of antiparallel interactions). Such  a situation is often encountered in
Mn$_2$YZ compounds, but not all of them exhibit non-collinearity. In general, these materials
crystallize in  the non-centro\-symmetric   ${I\overline{4}m2}$  structure
with  two non-equivalent Wyckoff  positions occupied by Mn atoms:  Mn$_{\,\rm I}$ at
$2b~(0,\nicefrac{1}{2},0)$ and Mn$_{\rm II}$ at $2d~(0,\nicefrac{1}{2},\nicefrac{3}{4})$.
Z and Y elements  occupy   the $2a~(0,0,0)$  and $2c~(0,\nicefrac{1}{2},\nicefrac{1}{4})$
positions, respectively~(Fig.~\ref{fig:jxc}\,a). 
The most significant exchange  coupling between the nearest Mn$_{\,\rm I}$
and Mn$_{\rm  II}$ atoms is  characterized by a large  exchange constant
(${J_{\rm Mn_I-Mn_{II}}\sim-20}$~meV) (e.g. \cite{MSR11}) that
leads to a typical collinear FiM
(ferrimagnetic) state. Despite the fact that the in-plane interaction of
Mn atoms can be rather complicated  (e.g. the nearest in-plane   neighbours   couple   parallel,   the   next-nearest   couple
antiparallel or parallel, and so on), these interactions are rather weak
compared to $J_{\rm  Mn_I-Mn_{II}}$, which always aligns  the Mn spin
moments of the same plane  parallel  to  one another   (Fig.~\ref{fig:jxc}\,b). For  this reason,  we initially do  not consider
the in-plane interactions but will expand the description in terms of the
effective {\it  inter-plane} exchange coupling $J$,  which indicates the
interaction of a certain Mn atom ($i$) with all other Mn atoms ($i'$) in
a different plane, i.e. ${J=\sum_{i'}J_{ii'}}$.

\indent
Since the collinear  order being substantially determined  by  the nearest-plane  $J$
interaction becomes even more stable if the Y atom is magnetic  (as
e.\,g. in case of Mn$_2$CoZ systems~\cite{MSR11}), our further
consideration concerns Mn$_2$YZ Heusler materials with the non-magnetic
heavy Y elements (such as Rh or Ir, since in case of light elements, such
as Ti or V, Mn atoms occupy equivalent $2c$ and $2d$ positions).
In this case, the collinearity can be perturbed by the next important interaction
$j$  between the next-nearest planes,  e.g. between  pairs  of Mn$_{\rm
  II}$-Y planes  as shown in Fig.~\ref{fig:jxc}\,b. This interaction is
 antiparallel due to its indirect origin realized through
 the main-group element Z (super-exchange)~\cite{LMM+13}.
Since  $j$ tends to rotate the moments of the nearest Mn$_{\rm II}$-Y planes antiparallel to each other, it
competes with the strong antiparallel  exchange $J$, and may then result
in a non-trivial canting angle (${\theta\ne0^\circ,180^\circ}$,
Fig.~\ref{fig:jxc}\,b).  The relevant  $\theta$-dependent part of the Heisenberg Hamiltonian will contain only antiparallel interactions:
\begin{equation}
  H_{\theta} = - J\cos\theta - \nicefrac{1}{2}\cdot j\cos2(\pi-\theta)\,,
\label{eq:heisenberg-model}
\end{equation}
where the first term is the coupling of the nearest planes (Mn$_{\,\rm I}$-Z with Mn$_{\rm II}$-Y) and the second is that of the next-nearest
(Mn$_{\rm II}$-Y) planes. The factor $\nicefrac{1}{2}$ accounts for
the twice sparser entrance of the next-nearest plane couplings. The extrema  of $H_{\theta}$ are found from:
\begin{equation}
  \sin\theta\left(\nicefrac{1}{2}+\frac{j}{J}\cos\theta\right)=0\,,
\end{equation}
and ${\theta_{1,2}=180^\circ\pm\arccos\left(\frac{J}{2j}\right)}$ non-collinear solutions
are given subject to the condition  $j/J>\nicefrac{1}{2}$, which means that the canting occurs only if the next-nearest antiparallel exchange $j$ is sufficiently strong.


\indent
To justify the proposed magnetic order we performed
{\it ab initio} calculations (Supporting Information~\cite{SuppInfo}, Sec.~V) for Mn$_2$RhSn and another two
similar Heusler systems, Mn$_2$PtIn and Mn$_2$IrSn.  For Mn$_2$RhSn, the
plot of the total energy as a function of $\theta$ indeed exhibits two energy minima corresponding to the non-trivial
canting angles $\theta_{1,2}=180^\circ\pm55^\circ$. Similar plots
were obtained for another two compounds (Supporting Information~\cite{SuppInfo}, Fig.~S8).
Calculated local moments, their orientations, total
magnetization, and  experimentally measured  one, are summarized in
Tab.~\ref{TAB:abinitio}.
\begin{table}
\begin{tabular}{l|c|c|c|c|c|c}
\hline
Mn$_2$YZ & $m_{\rm Mn_I}$ & $m_{\rm Mn_{\rm II}}$ & $\theta_{1,2}[^{\,\circ}]$ & $m_{\rm Y}$ & $M$ & $M_{\rm exp}$\\ \hline\hline
Mn$_2$RhSn & 3.51 & 3.08 & $180\pm55$  & 0.14 & 1.9 & 1.97\\ \hline
Mn$_2$PtIn & 3.38 & 3.30 & $180\pm50$  & 0.12 & 1.4 & 1.6\\ \hline
Mn$_2$IrSn & 3.52 & 3.08 & $180\pm44$  & 0.09 & 1.4 & 1.5\\ \hline
\end{tabular}
\caption{Computed atomic magnetic moments $m$,  canting   angles $\theta_{1,2}$ and
  total magnetization per formula unit ${M=m_{\rm Mn_I}+m_{\rm Mn_{\rm II}}\cdot\cos\theta+m_{\rm Y}}$, compared to the experimentally
  measured magnetization~$M_{\rm exp}$. Values of magnetic moments/magnetization are given in $\mu_{\rm B}$.
  \label{TAB:abinitio}}
\end{table}
These magnetic properties may be significantly affected by
those kinds of disorder which are typical for Heusler systems. The
details of this aspect are discussed in Supporting Information~\cite{SuppInfo} (Sec.~IV).

\begin{figure}
\centering
  \includegraphics[clip,width=0.8\linewidth]{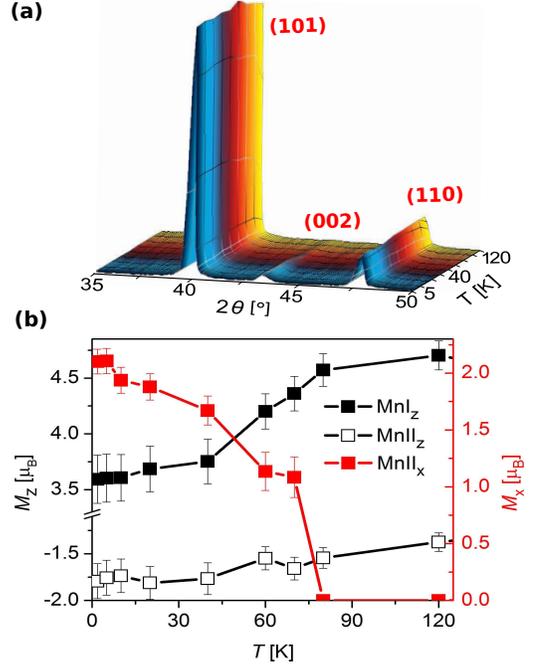}
  \caption{(a)~Temperature-dependent neutron diffraction spectra. The (002)-peak decays over 1.8-80~K.
    (b)~Weakening of the in-plane magnetism (produced by Mn$_{\rm II}$ $x$-component)
    releases the $z$-component of Mn$_{\,\rm I}$, while the $z$-component of Mn$_{\rm II}$
    evolves rather insignificantly.
\label{fig:1p8refined}}
\end{figure}

\indent
Powder  neutron scattering data  convincingly demonstrate  the predicted
gro\-und-state  non-collinearity (Fig.~\ref{fig:1p8refined}).
At 1.8~K the magnetic moments are 3.59 and 3.47~$\mu_{\rm B}$ on Mn$_{\,\rm I}$ and
Mn$_{\rm II}$. The value obtained for  the more localized
Mn$_{\,\rm I}$   correlates   with    the   calculated    result   (Tab.~\ref{TAB:abinitio}), while  the Mn$_{\rm II}$ moment  is larger: it
is defined less precisely as  the scattering events on itinerant moments
are more  dispersed.  The magnetic structure  was found to  be canted by
about  ${\theta_{1,2}=(180\pm58.9)^\circ}$  within alternating  Mn$_{\rm II}$-Rh planes. It is important to note, that such strong magnetic
canting was never reported for the Heusler materials, in which it is
typically of an order few degrees at most.

\begin{figure}
\centering
  \includegraphics[clip,width=1.0\linewidth]{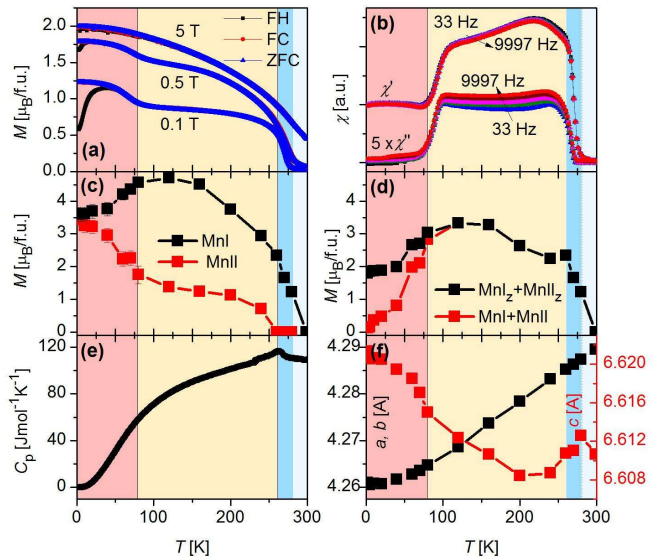}
  \caption{Evolution of the magnetic structure with the temperature. Canted (red), collinear ferrimagnetic (yellow) and disordered (blue) magnetic states of Mn$_2$RhSn. (a)~Zero-field-cooled~(ZFC), field-cooled (FC) and
    field-heated (FH) magnetization as a function of the temperature
    measured at induction fields of 0.1, 0.5, and 5~T. (b)~Real
    ($\chi'$) and imaginary ($\chi''$)  ac-susceptibility components
    are frequency independent and show a pronounced step at the
    onset of the FiM phase. (c)~The change in the canting
    angle occurs because of the simultaneous re-alignment of the
    Mn$_{\rm II}$ moment and a decrease in its absolute value. This, in turn,
    releases the previously suppressed  Mn$_{\,\rm I}$
    moment from 3.5 to 4.5 $\mu_{\rm B}$. (d)~The sum of the total and
    $z$-components of the Mn$_{\rm I(II)}$ moments follows the
    ac-susceptibility behavior. No in-plane component is present after
    80~K. (e)~A change in the slope of the heat capacity curve is
    observed in the vicinity of the spin-reorientation. (f)~Evolution of the lattice parameters with
    temperature: the change in the magnetism is echoed mainly by the
    $c$-parameter, while $a$ evolves monotonically.
\label{fig:t-dependency}}
\end{figure}

\indent
Being non-collinear in the ground state, the  magnetic
con\-fi\-gu\-ra\-tion  evolves  with  changes   to  the  temperature
and  external field.  Observation of  the  (002)-peak  intensity  for 
${T\leq80}$~K      indicates       the      presence      of      in-plane
magnetism~(Fig.~\ref{fig:1p8refined}\,b).  As the temperature increases,
the  peak gradually  decreases and  subsequently vanishes  for ${T>80}$~K,
suggesting   that  the   in-plane  magnetic   component   is  suppressed
(Fig.~\ref{fig:1p8refined}\,b).  This  is   attributed  to  the  gradual
spin-reorientation of  the  Mn$_{\rm II}$  sublattice;  the
canting angle  decreases until a collinear FiM  order sets in at
80~K.  Such  behavior  is  strongly  pronounced in  the  $M(T)$  curves
measured in  weak fields (0.1-0.5~T,  Fig.~\ref{fig:t-dependency}\,a)
and
suppressed  in stronger  fields (5~T).  This is  evidently  an intrinsic
effect as  the applied  fields are larger  than the  coercive field~(${H
  \rm_c=0.065~T}$).  It  is  not  only  the mutual  orientation  of  the
site-specific moments that changes but their absolute values also change
(Fig.~\ref{fig:t-dependency}\,c). In the canted lowest-temperature state,
the Mn$_{\,\rm  I}$ moment is  somewhat compensated by the  equally strong
Mn$_{\rm  II}$. As the  temperature increases,  the moments  of Mn$_{\rm
  II}$  delocalize  further  and  release  the  Mn$_{\,\rm  I}$  to  reach
4.5~$\mu_{\rm B}$. This  occurs gradually, and the  slope of the
zero-field heat  capacity curve changes (Fig.~\ref{fig:t-dependency}\,e);
the spin-wave term is sufficiently  weak in comparison to the electronic
and phonon contributions that no  sharp anomaly is visible. However, the
onset of  the FiM phase  is characterized by  the explicit
step-like      increase     in     the      ac-susceptibility     signal
(Fig.~\ref{fig:t-dependency}\,b). Measured values of $\chi'$ and $\chi''$
were  found  to be  independent  of  the  frequency, suggesting  a  high
magnetic homogeneity. The evolution of the magnetism with temperature is
echoed           by            the           crystal           structure
(Fig.~\ref{fig:t-dependency}\,f).  Although  the $a$-parameter  increases
monotonically, the change in $c$-parameter is non-linear and corresponds
to the ac-susceptibility  behavior. The sudden rise in  the vicinity of
280~K  is an  anomaly corresponding  to $T_{\rm  C}$.  The $c$-parameter
eventually decreases  until a  transition to the  cubic phase  occurs at
about $570~^\circ$C (Supporting Information~\cite{SuppInfo}, Fig.~S5).

\indent
{By systematic coarse-graining of the spin-lattice
model a micromagnetic continuum theory has been developed
(Supporting Information~\cite{SuppInfo}, Sec.~VI).
Considering only the leading Heisenberg-like exchange,
the analysis shows that in tetragonal inverse
Mn$_2$YZ Heusler alloys, the magnetic ordering  displays coexisting magnetic modes
with ferrimagnetism (FiM) of the two sublattices and an
antiferromagnetic mode (AFM) on the Mn$_{\rm II}$-sublattice.
These systems, thus, are close to a bicritical (or tetracritical) point
in their magnetic phase diagram.
In Mn$_2$RhSn, the thermodynamic potential favours a dominating
collinear FiM order for ${T>80}$~K. Below this temperature
the AFM sets in.
%
%
By the crystal symmetry of Mn$_2$YZ, \textit{ chiral inhomogeneous DM couplings}
exist in spatial directions perpendicular to the crystal axis \cite{BY89,BRWM02}
that cause a spiral twist of these magnetic modes with long pitch.
The micromagnetic model for the FiM state
is exactly the Dzyaloshinskii-model for a magnetic
order in acentric tetragonal crystals
from $\overline{4}2m$ ($D_{2d}$) class~\cite{Dz64,BY89}.
Therefore, in the collinear FiM state, chiral skyrmions and skyrmion
lattices exist in these magnets, as predicted in Ref.~\cite{BY89,BH94}.
The micromagnetic model predicts a chiral
twisting length ${\Lambda\sim130}$~nm, which corresponds
to the diameter of the FiM-state skyrmions.
These chiral skyrmion states exist in the inverse Heusler alloys
without the need of any additional effects not accounted for
by the basic magnetic couplings, i.e.  Heisenberg-like
and DM-exchange and leading anisotropies, and
at arbitrary temperature.
This is in contrast to chiral cubic helimagnets, which
require fine-tuned additional effects for the existence
of skyrmionic states.
Because the tetragonal lattice also induces a
sizeable easy-axis magneto-crystalline anisotropy in Mn$_2$RhSn,
as calculated by relativistic DFT, the magnetic phase diagram
is not expected to display a field-driven condensed skyrmion phase
in this FiM-state.
The ratio of easy-axis anisotropy to DM coupling is large.
Using the universal phase diagram of chiral magnets~\cite{BH94},
skyrmions do exist as nonlinear solitonic excitations
of the collinear state in Mn$_2$RhSn.
Therefore, this inverse Heusler alloy is an ideal system
to realize reconfigurable nanomagnetic patterns composed
of its two-dimensional free skyrmions at elevated temperatures.

Coexistence of FiM and AFM orders in the canted state
will be the subject to different DM-couplings. Thus, in
the ground-state, novel types of chirally twisted textures
can exist in Mn$_2$YZ alloys. These chiral DM-couplings
favour different twisting lengths for the two modes
and their interaction. Then, the magnetic order may become quasiperiodic and
exceedingly complex, as recently described for the similar
case of textures in biaxial nematic liquid crystals~\cite{GoloKats13}.
The presence of several DM-terms and anisotropies
affecting the coexisting magnetic modes promise a rich behavior
of chiral textures in tetragonal inverse Heusler alloys Mn$_2$YZ.
E.g., closely below the onset of spin-reorientation temperature,
the chiral skyrmion of the FiM-state is superimposed by
a vortex-like AFM-configuration on the Mn$_{\rm II}$-sublattice
with a defect in the core of the soliton configuration.
In the ground-state, such configurations may become
instable, depending on the stiffness of the AFM order.
Up to now, such complex configurations have been analysed
only for the simpler case of chiral AFMs
with a coexisting weak-FM mode \cite{BRWM02}.

\indent
As we have demonstrated theoretically and experimentally,
the  design  of  non-collinear  magnets  within the  Heusler  family  of
materials can  be based on Mn$_2$YZ  compositions, with Y and  Z being a
non-  or  weakly-magnetic  transition-metal  and a  main-group  element,
respectively.  The  choice of  the  Mn$_2$YZ  Heusler  group allows  to
control the canting angles by, e.g. combining
the  Y and  Z elements  or varying  the Mn  content.  The  use  of heavy
transition metals (e.g. as in the present case, Y=Pt, Rh, Ir and Z=Sn,
In)  amplifies the  magnetically-relevant  relativistic effects
that   are   already   present   in   these   systems,   such   as   the
DM interaction  and magneto-crystalline  anisotropy.  Such multiple
magnetic  degrees of  freedom together  with the
possibility  of their manipulation  provided by  the family  of Mn$_2$YZ
Heusler materials  is vital for efficient  engineering and stabilization
of  various magnetic  orders.  In particular, Mn$_2$RhSn is suggested to
be a promising candidate for realizing the  skyrmion state in the
Heusler family.

\acknowledgments
Authors thank A. Bogdanov, R.~Stinshoff and A.~Beleanu for helpful discussions. The support of the European Commission under
the 7th Framework Programme: Integrated Infrastructure Initiative for Neutron Scattering and Muon Spectroscopy: NMI3-II/FP7 -
Contract~No.~283883, the Graduate School of Excellence
'Materials Science in Mainz', the  DFG project FOR~1464 'ASPIMATT'
and the European Research Council (ERC) for 'Idea Heusler!' are gratefully acknowledged.

\end{document}


\title{Supplemental: Large Noncollinearity and Spin Reorientation in Novel Mn$_2$RhSn Heusler Magnet}

\author{O.~Meshcheriakova,$^{1,2}$ S.~Chadov,$^2$ A.~K.~Nayak,$^2$ U.~K.~R\"o\ss ler,$^{3}$ J.~K\"ubler,$^4$ G.~Andr\'{e},$^5$ A.~A.~Tsirlin,$^2$ J.~Kiss,$^2$ S.~Hausdorf,$^2$ A.~Kalache,$^2$
  W.~Schnelle,$^2$ M.~Nicklas,$^2$ and C.~Felser$^2$}
\affiliation{$^1$Graduate~School~of~Excellence~''Materials~Science~in~Mainz'' Johannes Gutenberg - Universtit\"{a}t,  55099 Mainz, Germany}
\affiliation{$^2$Max-Planck-Institut f\"{u}r Chemische Physik fester Stoffe,  N\"othnitzer Strasse~40, 01187 Dresden, Germany}
\affiliation{$^3$Leibniz-Institut f\"{u}r Festk\"orper- und  Werkstoffforschung IFW, Helmholtz Strasse~20, 01171 Dresden, Germany}
\affiliation{$^4$Institut f\"{u}r Festk\"{o}rperphysik, Technische Universit\"{a}t Darmstadt, 64289 Darmstadt, Germany}
\affiliation{$^5$Laboratoire Léon Brillouin, CEA-CNRS Saclay, 91191 Gif-sur-Yvette Cedex, France}

\maketitle

\section{Synthesis of \boldmath ${\rm Mn_2RhSn}$ and ${\rm Mn_2IrSn}$}
Polycrystalline samples were repeatedly arcmelted from
stoichiometric amounts of high-purity commercially available
elements in an Ar atmosphere with an overall mass loss of less than
0.5~wt.\%. In each melting run, a piece of titanium was used to
purify the residual atmosphere. A two-step process was employed for
Mn$_2$RhSn: a premelt of Rh-Sn was prepared, and this was then
placed on the Mn. According to the respective phase diagrams, Rh and Sn react well
with each other and form a stable phase. In the second step, when
the premelt is heated, it absorbs the Mn pieces, and the evaporation of Mn is
minimized. To ensure homogeneity, the samples were melted 3 times on
each side. As a result, after the Mn is absorbed by the phase,
evaporation of the complete phase, not the single elements, takes
place.

The Mn$_2$IrSn sample was prepared by induction heating. The
procedure was repeated several times to ensure homogeneity: six
repetitions of arcmelting and two of induction heat. In the latter
process, the sample was maintained in the liquid state for 5 min.
After 1 week of annealing, the arcmelted samples were fast cooled,
whereas the induction-heated samples were cooled slowly. The ingots
were then wrapped in Ta foil and annealed in evacuated silica tubes
at 800$^{\circ}$C for 1 week. To reduce the amount of surface
oxidation, Mn pieces were preliminary sealed in evacuated silica
tubes and left overnight at 900$^\circ$C for purification. These
pieces were processed repeatedly until a shiny silver-coloured
surface was obtained.

\section{Prior characterization}
\begin{figure}
\centering
  \includegraphics[clip,width=0.9\linewidth]{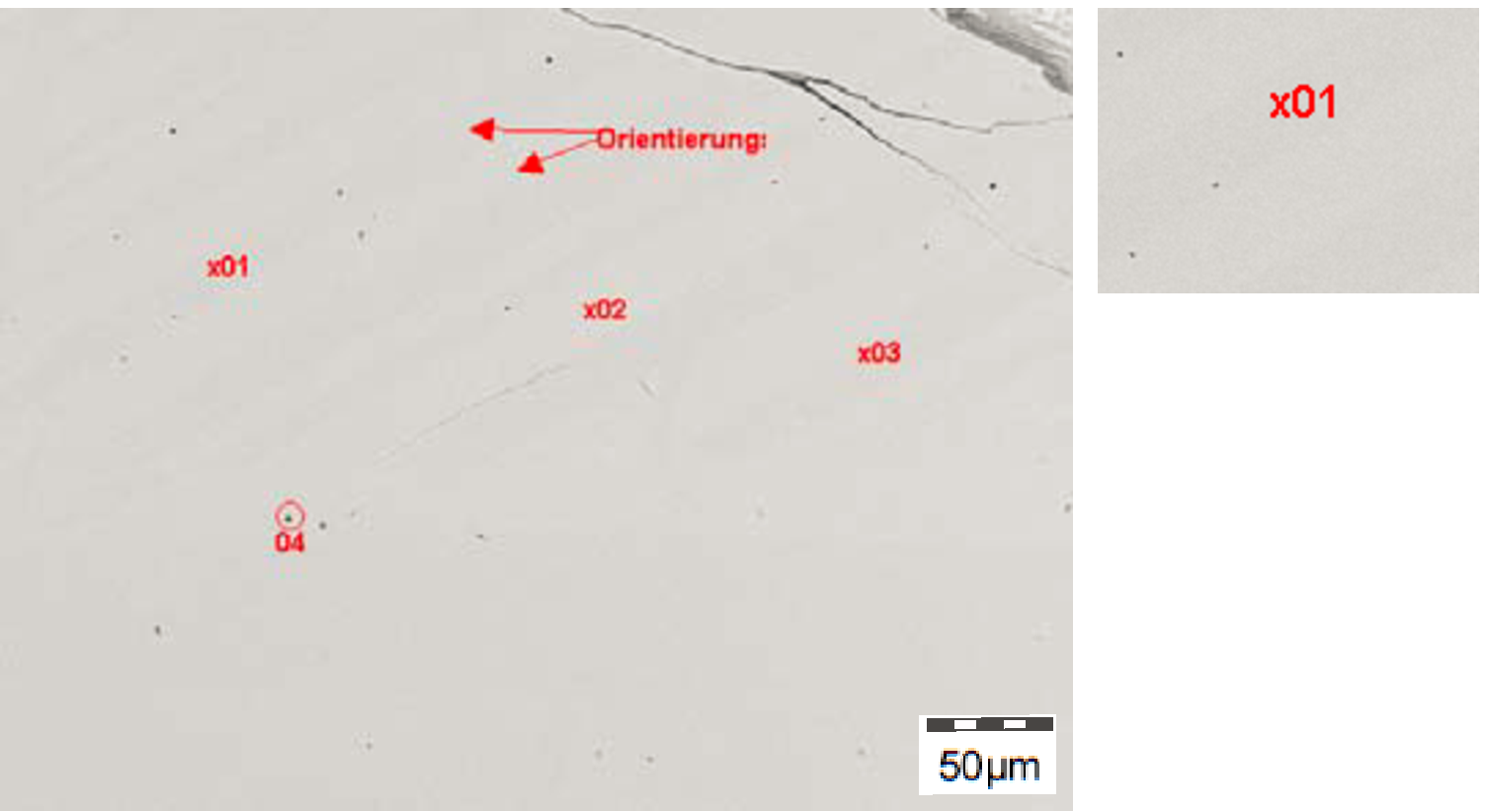}
\caption{SEM image of Mn$_2$RhSn; composition
analysis was performed at the areas marked. The Mn$_2$RhSn
stoichiometry is constant across the whole sample; a minor impurity
of Mn$_{3.5}$RhSn is present at around 1 at.$\%$ and does not bias
the presented results.
\label{fig:SEM}}
\end{figure}

\begin{figure}
\centering
  \includegraphics[clip,width=0.9\linewidth]{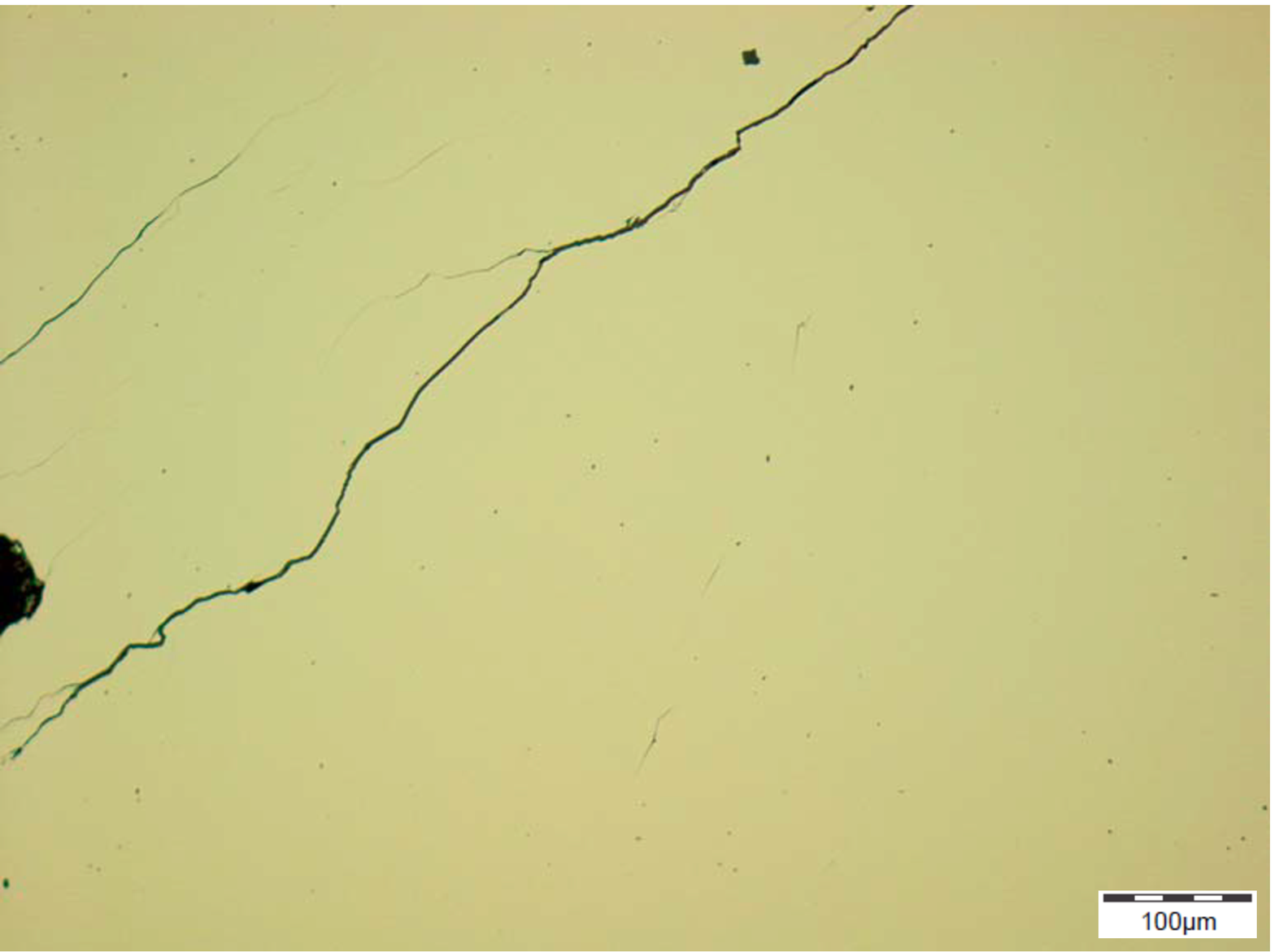}
  \caption{{An optical microscope image of the homogeneous Mn$_2$RhSn phase.}
\label{fig:OM}}
\end{figure}

Metallographic analysis by scanning electron (Fig.~\ref{fig:SEM})
and optical (Fig.~\ref{fig:OM})
microscopy revealed that the samples are single-phase materials with
a homogeneous composition distribution. The composition was
characterized by energy-dispersive X-ray (EDX) spectroscopy (values
are summarized in Tab.~\ref{tab:EDX}).
\begin{table}[htb]
\centering
\begin{tabular}{cccc}
\hline Spot & Mn (at.\%)& Rh (at.\%)& Sn (at.\%)\\
\hline
1) & 49.73 & 25.55 & 24.71 \\
2) & 48.79 & 25.73 & 25.48 \\
3) & 47.84 & 26.23 & 25.93 \\ \hline
\end{tabular}
\caption{EDX analysis of the Mn$_2$RhSn sample taken from the areas
indicated in Fig.~\ref{fig:SEM}. The composition is
well reproduced across the whole observed area.\label{tab:EDX}}
\end{table}
Since the electron penetration depth is on the order of nanometres, a
well-polished sample surface is essential for eliminating morphology
effects. For this reason, the samples were embedded in epoxy resin
blocks, and a smooth surface was prepared. The measured composition
deviates from the target values by 0.5~at.\%, which is within the
range of experimental error.

\subsection{X-ray diffraction}
\begin{figure}
  \centering
  \includegraphics[width=1.0\linewidth,clip]{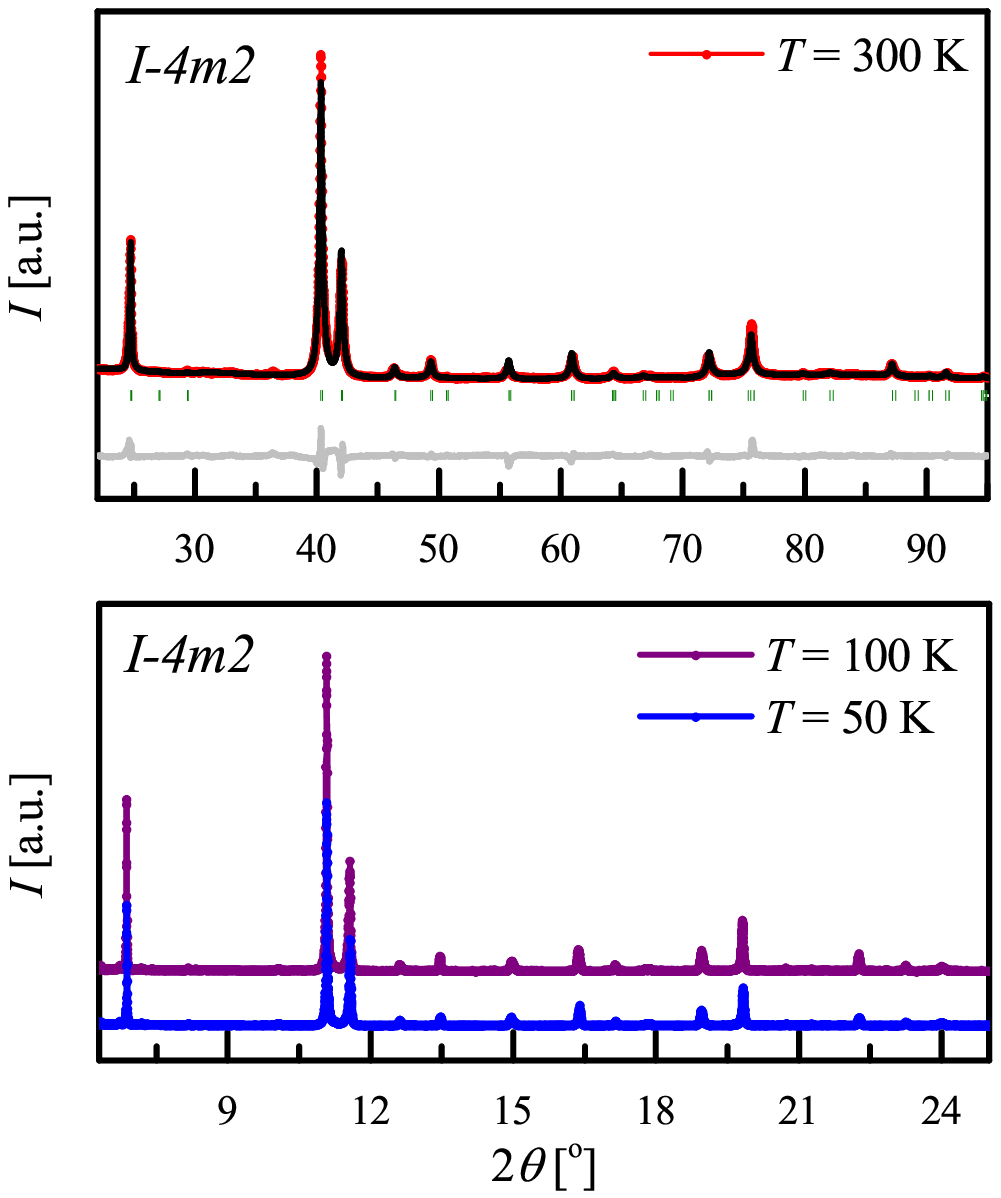}
\caption{Powder X-ray patterns obtained at room temperature (top)
and 50 and 100~K (bottom). The coloured and black lines correspond
to the observed and calculated intensities, respectively. Incident
light wavelengths of ${\lambda=1.5405}$~\AA\ and 0.43046~\AA\ were
used for the room- and low-temperature measurements, respectively.
The high-resolution XRD data are consistent with the $I$-$4m2$ symmetry
and confirm the tetragonal crystal structure at low temperatures.
\label{fig:xrd_rt_100_50}}
\end{figure}

Powder X-ray diffraction patterns (Fig.~\ref{fig:xrd_rt_100_50})
were obtained using a Guinier camera
(Cu~K$_\alpha$ radiation) with LaB$_6$ acting as the internal
standard. The samples were first sieved through a 40~$\mu$m mesh.
Rietveld refinement was performed using the FullProf
software~\cite{FULLPROF04} for the structure analysis.

In the case of Mn$_2$RhSn~\cite{AMW+13}, low-temperature X-ray
diffraction patterns were measured at ESRF, Grenoble, with an
incident beam wavelength of 0.43046~\AA\ for a high-resolution data
analysis. The Mn$_2$RhSn compound crystallizes in an inverse
tetragonal 119 Heusler structure (with Wyckoff sites of Mn$_{\rm
I}$: (0, 0, $\frac{1}{2}$); Mn$_{\rm II}$: (0, $\frac{1}{2}$,
$\frac{1}{4}$); Rh: (0, $\frac{1}{2}$, $\frac{3}{4}$); and Sn: (0,
0, 0)). Mn$_2$IrSn and Mn$_2$PtIn~\cite{NSW+12} crystallize in the same structure with lattice parameters of
${a=4.29}$~\AA, ${c=6.59}$~\AA\ and ${a=4.32}$~\AA, ${c=6.77}$~\AA, respectively.

\subsection{Magnetic and thermal measurements}
\begin{figure}
\centering
  \includegraphics[clip,width=0.9\linewidth]{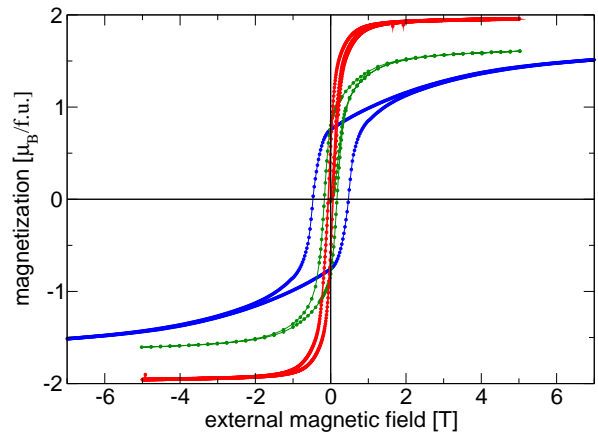}
  \caption{{Magnetic hysteresis loops measured at 1.8~K of
      polycrystalline Mn$_2$RhSn~\cite{AMW+13} (red), Mn$_2$PtIn~\cite{NSW+12} (green),
      and Mn$_2$IrSn (blue; present work) samples.}
\label{fig:hysteresis}}
\end{figure}

\begin{figure}
  \centering
  \includegraphics[clip,width=0.9\linewidth]{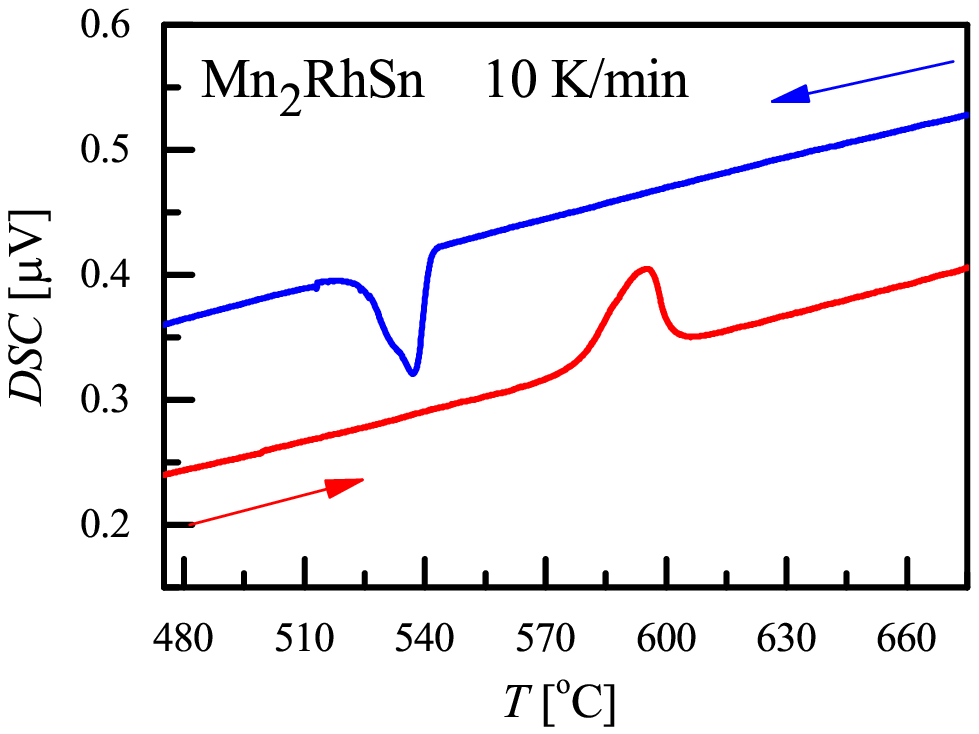}
\caption{The transition from the high-temperature cubic to
low-temperature tetragonal phase occurs between 537 and
594$^\circ$C. Red and blue curves correspond to the heating and cooling regimes, respectively. \label{fig:cubic-phase-transition}}
\end{figure}

Magnetization measurements were performed in constant field sweeps
at different temperatures using the Quantum Design MPMS XL
superconducting quantum interference device (SQUID) magnetometer.
The total magnetization was obtained from the hysteresis loop at
1.8~K (Fig.~\ref{fig:hysteresis})
and was
1.97~$\mu_{\rm B}$ (per formula unit). In the zero-field-cooled
(ZFC) mode, the sample was initially cooled in the absence of a
field down to 2\,K,  and data were collected as the temperature was
increased in the applied field. In the field-cooled (FC) mode, data
were collected while the sample was cooled in the field, and
subsequently, data were also collected while the sample was heated
in the field during the field-heated (FH) mode. The real ($\chi'$)
and imaginary ($\chi''$) parts of the ac-susceptibility were
obtained simultaneously at the lowest possible dc-field of 50~Oe.
Various field frequencies from 33 to 9997~GHz were applied over the
temperature range of 2 to 300~K. The heat capacity measurements were
performed in zero field over the same temperature range. A
transition to the high-temperature cubic phase (Fig.~\ref{fig:cubic-phase-transition})
was observed with the help of
differential scanning calorimetry (DSC) at a moderate rate of
10~K/min; the powder sample was encapsulated in an Al$_2$O$_3$
crucible and measured in an Ar atmosphere.

\section{Neutron scattering}
A two-axis diffractometer equipped with a vertical focusing
pyrolytic graphite monochromator and a cold neutron guide was used
for the neutron scattering measurements. The sample was encapsulated
in a vanadium crucible, and the sample environment was constant
throughout the measurement. The magnetic and nuclear structures were
refined by the Rietveld method using the FullProf
software~\cite{FULLPROF04}. The nuclear phase was first optimized
above the magnetic ordering temperature, and the obtained parameters
were later used for the refinement of the magnetic phase. This
significantly improved the sigma deviations and $R$ factors. The
background was modelled by interpolation between manually selected
points. The peak shape profile was described by a pseudo-Voigt
function with a refined ratio between the Gaussian and Lorentzian
contributions. Different oxidation states, Mn$_{\rm I}^{3+}$ and
Mn$_{\rm II}^{2+}$, were assumed to calculate the magnetic
scattering form factors. The results of the neutron scattering
measurements are listed in  Tab.~\ref{tab:ref}.
\begin{table}
\centering
\begin{tabular}{ccccccc}
\hline Mn$_{\rm I}$~[$\mu \rm _B$] & \multicolumn{3}{c}{Mn$_{\rm
II}$~[$\mu
     \rm _B$]} & $\theta$ [$^\circ$] &  $a$~[\AA] & $c$[\AA]  \\ \cline{1-4}
${M_{\rm z}=M_{\rm tot}}$ & $M_{\rm x}$ & $M_{\rm z}$ & $M_{\rm tot}$ &
& & \\
\hline
3.59 & 2.10 & $-$1.80 & 3.47 & 58.9 & 4.261 & 6.261  \\
\hline
\end{tabular}
\caption{Refined values of the powder Mn$_2$RhSn sample obtained
from neutron scattering measurements.\label{tab:ref}}
\end{table}

Several refinement approaches were used by fixing or releasing
certain, all, or some parameters. In total, we varied 10 parameters:
the scale, zero-shift, $a$-constant, $c$-constant, projections of
the spin [$M_{\rm z}$ (Mn$_{\rm I}$), $M_{\rm x}$ (Mn$_{\rm II}$),
and $M_{\rm z}$ (Mn$_{\rm II}$)], Lorenztian-to-Gaussian ratio in
the peak shape (varied separately for the Bragg and magnetic phase),
and overall isotropic displacement (temperature) factor. The
room-temperature pattern was refined first to eliminate any
contribution of the magnetic signal, and the obtained zero-shift
value was then fixed to avoid errors in the lattice parameters.

The lattice constants are robust and independent of the specific
refinement procedure. We also found that the Mn$_{\rm I}$ moment was
indeed strongly localized and aligned along the crystallographic
$z$-axis. If a slight deviation from this direction is introduced,
then the refinement does not converge. In contrast, the Mn$_{\rm
II}$ moment prefers a canted orientation, and the obtained absolute
value of the Mn$_{\rm II}$ moment was slightly higher than that
predicted theoretically: 3.47~$\mu_{\rm B}$ as opposed to
3.08~$\mu_{\rm B}$. The value of the Mn$_{\rm I}$ moment, however,
was in good agreement with the predicted value: 3.59 $\mu \rm _B$ to
3.51 $\mu \rm _B$.

\begin{figure}
  \centering
  \includegraphics[width=0.9\linewidth,clip]{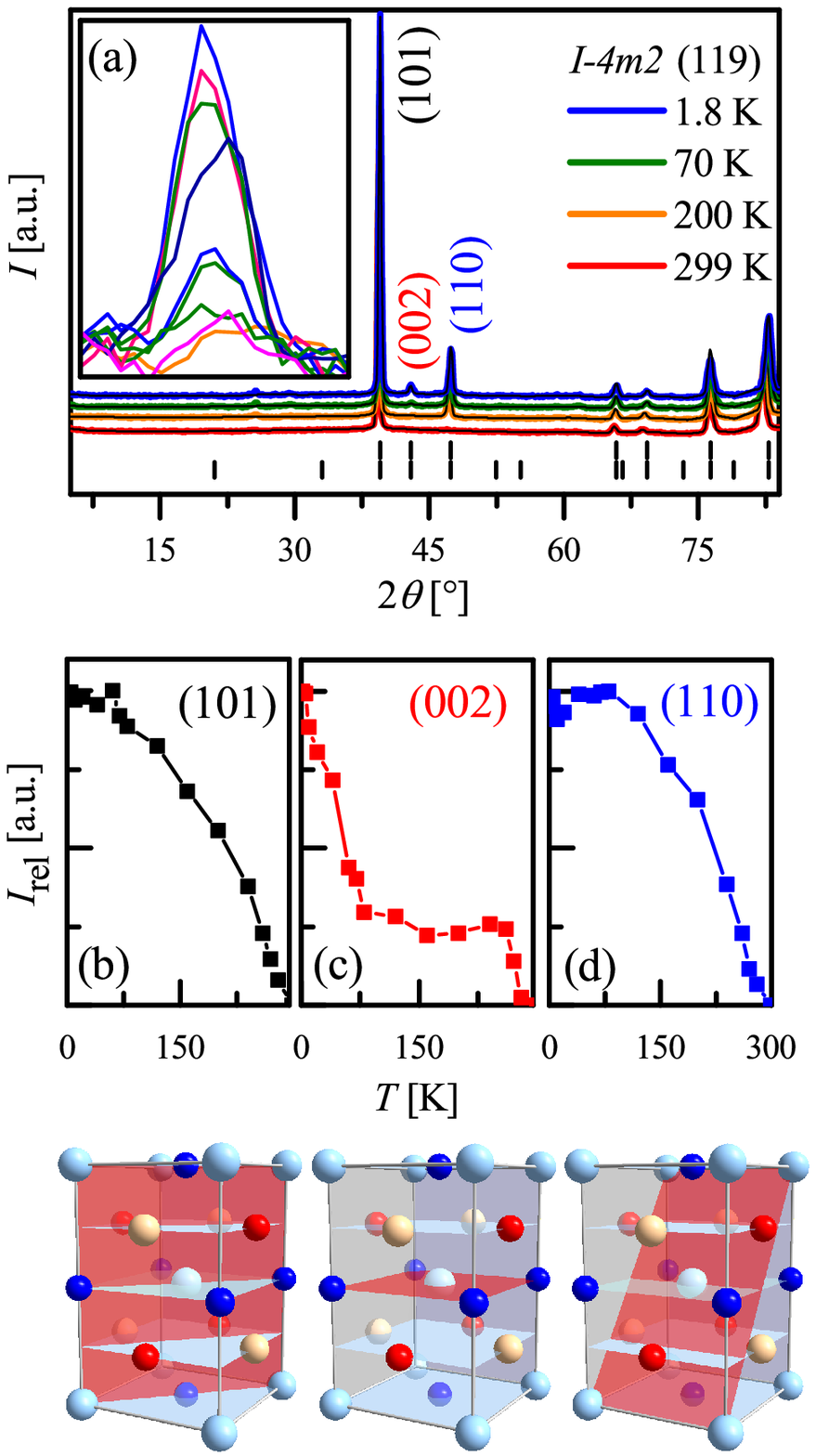}
  \caption{Neutron data of Mn$_2$RhSn. (a)~Evolution of the neutron
    spectra with temperature (1.8~K, 70~K, 200~K and 299~K); peak
    positions from the Bragg and magnetic phases are distinguished by
    vertical markers. Inset: evolution of (002) peak within the whole
    temperature regime. Normalized intensity of the (101), (002) and
    (110) peaks ((b), (c), (d) panels, respectively) together with the corresponding crystal planes.
      \label{fig:npd1}}
\end{figure}

The magnetic state appears to be well-analyzable due to its noticeable contribution to the overall intensity: e.g., (101)-peak increases by
nearly 65\,\% with temperature decrease from ${T=299}$~K (paramagnetic) to 1.8~K. The additional
intensity contains only that magnetic component, which is perpendicular to the scattering vector;
therefore the neutrons reflected from the (101), (002) and (110) crystal planes give us an estimate
how the magnetic moment evolves within these planes (see
Fig.~\ref{fig:npd1}). For ${T>80}$~K the magnetic contribution to the (002) peak vanishes and only the Bragg intensity is
observed. The in-plane magnetism is realized by canting of the Mn$_{\rm II}$ moment.

\section{\boldmath Disorder effects and the phase stability of ${\rm Mn_2RhSn}$}
\label{sec:stability}
Chemical disorder, which often occurs in multicomponent
  systems, such as Heusler alloys, may severely influence the magnetic
  properties. Random exchange between Rh and Mn$_{\rm I}$ would increase
  the amount of Mn$_{\rm II}$  type (magnetically antiparallel to
  Mn$_{\rm I}$). In turn, the exchange between Sn and Mn$_{\rm II}$
  will increase the amount of Mn$_{\rm I}$ type. 
The exchange between Rh and Sn will locally convert Mn$_{\rm I}$ into Mn$_{\rm II}$ within Mn$_{\rm I}$--Sn planes, and Mn$_{\rm II}$ into Mn$_{\rm I}$ – within Mn$_{\rm II}$--Rh planes. In which way particular type of disorder will affect the system exactly, is rather
complicated question, as it implies not just a straightforward redistribution of different Mn types,
but the change of the whole magnetic coupling picture. It is easy to show (e.g., by first-principle
calculations) that in all cases the total energy drastically increases, which indicates that such
events are of small probability (once the system holds the correct stoichiometry and is properly
annealed). In any case, if such situation would occur, the interpretation of the neutron spectra
using the proposed magnetic picture would be unreliable.

\begin{figure}
  \centering
  \includegraphics[width=\linewidth,clip]{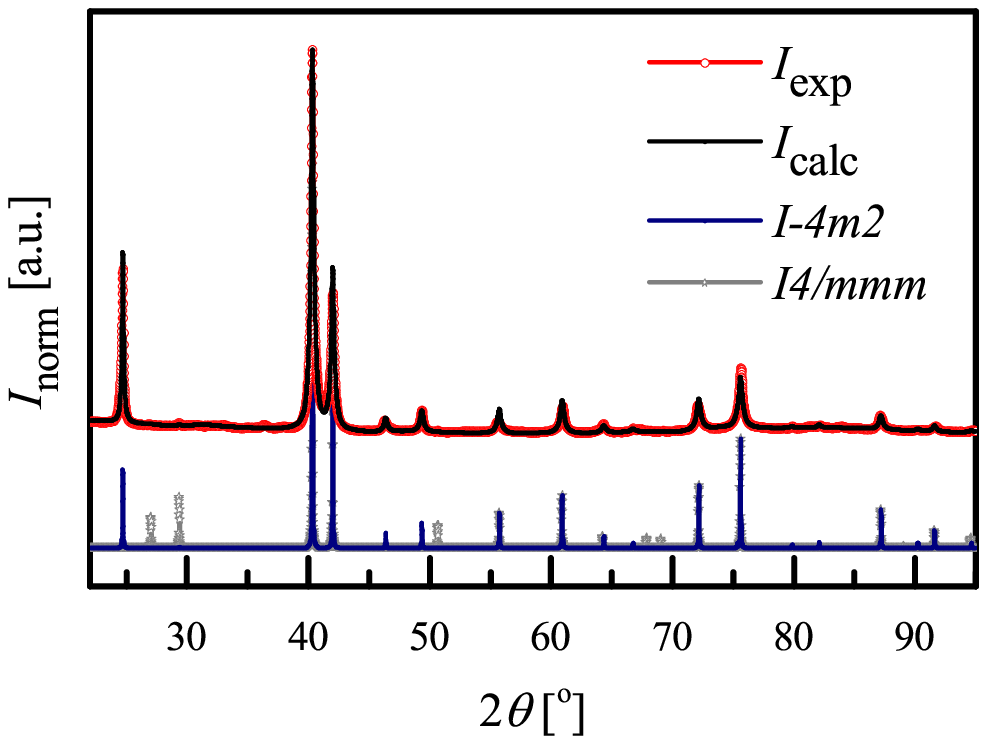}
  \caption{XRD powder pattern (red) of Mn$_2$RhSn. The measured
pattern has been refined using the noncentrosymmetric tetragonal Heusler
structure ${I{\textrm-}4m2}$ (No.~119) as a model. For comparison, both ${I{\textrm-}4m2}$ (119) and
${I4/mmm}$ (139) structures are showed below (blue and gray, respectively).
      \label{fig:119139}}
\end{figure}

For the present samples such straightforward reason for chemical
disorder as deviation between the actual and the target compositions can
be excluded due to the high quality evidenced by EDX and also XRD (see Fig.~\ref{fig:OM}, Tab.~\ref{tab:EDX}). Within this restriction, certain
random intermixing of different atomic types would be still possible. However, any mixtures involving Sn are unlikely, as it affects the zinc-blende sub-structure, which is the
``skeleton'' of any Heusler alloy (as one can represent the Heusler
system as zinc-blende plus extra transition element).
Certain intermixing within Rh--Mn$_{\rm II}$ plane would still be possible
(indeed, there are Heusler systems stabilized by such mechanisms,
e.g. Fe$_2$CuGa which exhibits Fe-Cu intermixing~\cite{KCF+13}). However
such disorder at high rates would lead to the statistical emergence of
the inversion symmetry, which contradicts to the present XRD data (Fig.~\ref{fig:119139}),
convincingly deducing the noncentrosymmetric group ${I{\textrm-}4m2}$ (No.~119) instead of
${I4/mmm}$ (No.~139).

Experimentally the compositional stability of the compound was investigated by
considering Mn- and Rh-excess regimes. The Mn-excess series
(Mn$_{3-x}$Rh$_x$Sn, with the composition step of 0.1)
always exhibit the phase separation in a form of a growing amount of Mn-rich Mn$_3$Sn phase
(group No.~194, hexagonal) and the ``host'' phase of Mn$_2$RhSn.
There is no direct transition between these crystal structures for the symmetry reasons: right at Mn$_{2.1}$Rh$_{0.9}$Sn
composition  the phase separation sets in, as observed with the help of
XRD, EDX, optical and electron microscopy.
Introduction of additional Rh in Mn$_2$RhSn system reduces the $c/a$ ratio and gradually brings the structure to the cubic phase. The smallest Rh content, which is enough to form a cubic structure is Mn$_2$Rh$_{1.17}$Sn. Thus, the Mn$_2$RhSn phase is rather sensitive to a slight stoichiometric deviation of Rh or Mn.
For the working composition (i.e., Mn$_2$RhSn) the Rietveld refinement of the 2:1:1 sample shows the $R$-Bragg factor of 3.998. If present, the disorder between Mn and Rh atoms would contribute additional intensities to (002), (110), (202) and (310) peaks whereas the (101), (103), (211),
(301), (321) and (215) would be suppressed (see Fig.~\ref{fig:119139})

\section{Computational details}
\label{sec:computational}

To justify the proposed magnetic order, first-principles calculations
using the spin-polarized relativistic Korringa--Kohn--Rostoker
(SPR-KKR) Green's function method~\cite{EKM11} within a local density
approximation~\cite{VWN80} were performed for several Mn$_2YZ$
Heusler materials: the recently synthesized Mn$_2$PtIn~\cite{NSW+12},
Mn$_2$RhSn~\cite{AMW+13}, and Mn$_2$IrSn, which is reported for the
first time in the present work. To determine the magnetic ground
states, we started with the experimental lattice parameters and
allowed for the self-consistent determination of local moments
including their amplitudes, directions, and periodicity.

\begin{figure}
  \centering
  \includegraphics[clip,width=0.85\linewidth]{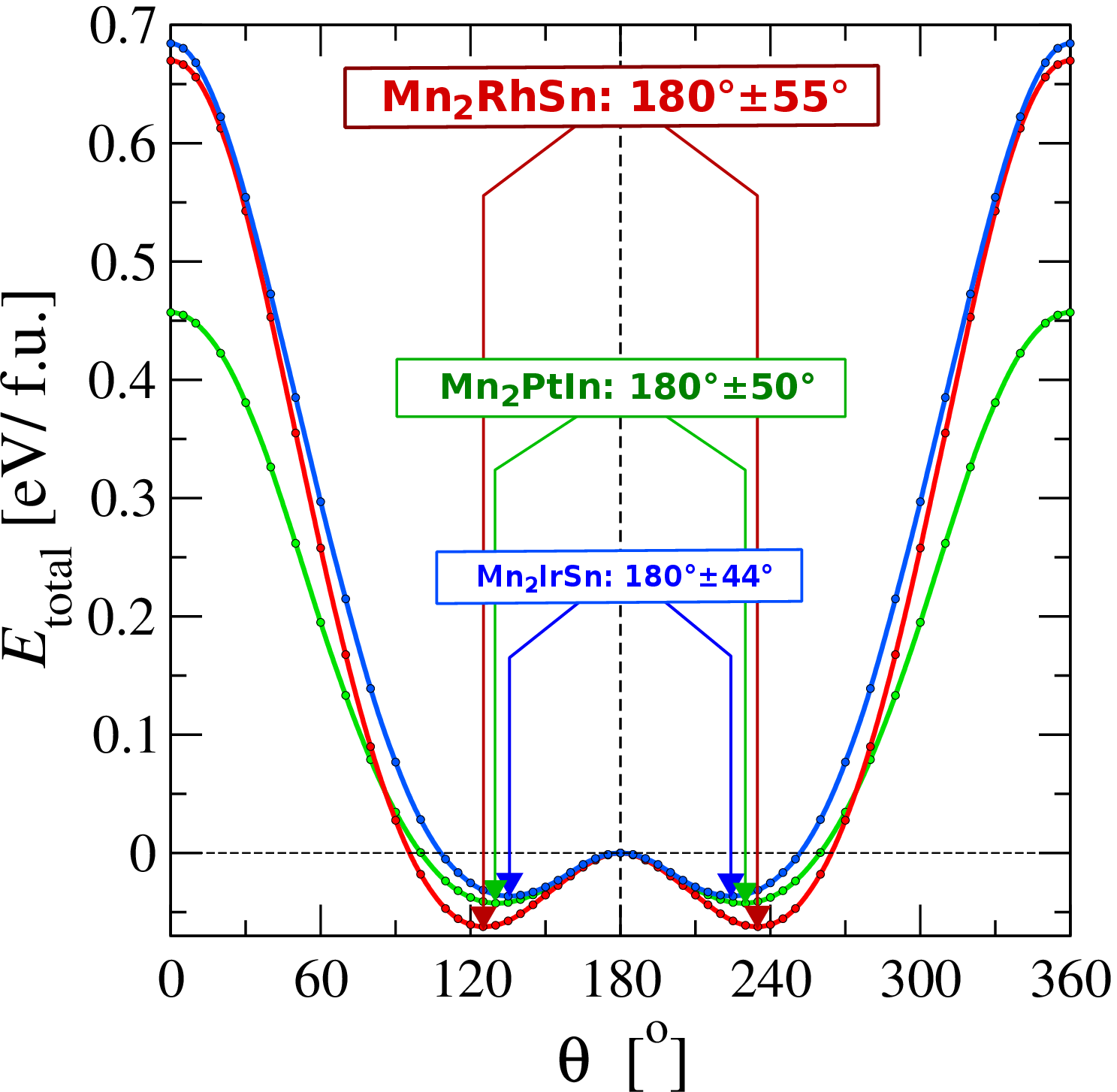}
  \caption{Total energy per formula unit computed as a function of
    the orientation of the Mn$_{\rm II}$ magnetic moment characterized by
    angle~$\theta$. The energy minima (indicated by arrows) occur at
    ${\theta_{1,2}=180^{\circ}\pm55^{\circ}}$, ${180^{\circ}\pm50^{\circ}}$,
    and ${180^{\circ}\pm44^{\circ}}$ in the case of Mn$_2$RhSn, Mn$_2$PtIn, and Mn$_2$IrSn, indicated respectively by red, green, and blue.
\label{fig:etotal_theta}}
\end{figure}

From the total energies obtained as a function of $\theta$~(Fig.~\ref{fig:etotal_theta}), not only both the Sn-containing
compounds but also the In-containing compound exhibit noncollinear
magnetic order characterized by canting of the Mn$_{\rm
  II}$ local moment direction: $\theta_{1,2}=180^\circ\pm55^\circ$,
$180^\circ\pm50^\circ$, and $180^\circ\pm44^\circ$ for
Mn$_2$RhSn, Mn$_2$PtIn and Mn$_2$IrSn, respectively. Upon closer
examination of Fig.~\ref{fig:etotal_theta},
the largest energy scale ($\sim$0.67, 0.45, and 0.6~eV for Mn$_2$RhSn,
Mn$_2$PtIn, and Mn$_2$IrSn, respectively) is revealed to be the
difference between two collinear configurations: the FM (ferromagnetic,
${\theta=0^{\circ}}$) and FiM (ferrimagnetic, ${\theta=180^{\circ}}$)
configuration. Thus, the small energy (0.06, 0.04, and 0.09~eV
for Mn$_2$RhSn, Mn$_2$PtIn, and Mn$_2$IrSn, respectively) gained by
canting can be considered as a perturbation of the collinear
ferrimagnetic state, which is typical for most Mn$_2$-based Heusler
systems.

To ensure that the canted magnetic state that is obtained is due to
the proposed mechanism, we computed the exchange coupling constants
(using the approach in Ref.~\cite{LKAG87}) for the model~Eq.~(1,~3) and
calculated the canting directly from minimizing the Heisenberg
Hamiltonian~(1). For example, the values obtained for Mn$_2$RhSn
(${J\!=\!-63.46}$ and ${j\!=\!-53.05}$~meV) satisfy criterion (2):
${j/J=0.83>\nicefrac{1}{2}}$. This then leads to
${\theta_{1,2}=180^{\circ}\pm53.3^\circ}$, which is in reasonable
agreement with the {\it ab initio} calculated value in Tab.~I (in the main part). The same
holds also for the In-based compound Mn$_2$PtIn: ${J=-36.64}$ and
${j=-27.67}$~meV gives ${j/J=0.75>\nicefrac{1}{2}}$, leading to
${\theta_{1,2}=180^{\circ}\pm48.6^\circ}$.

\begin{figure}
  \centering
  \includegraphics[clip,width=0.95\linewidth]{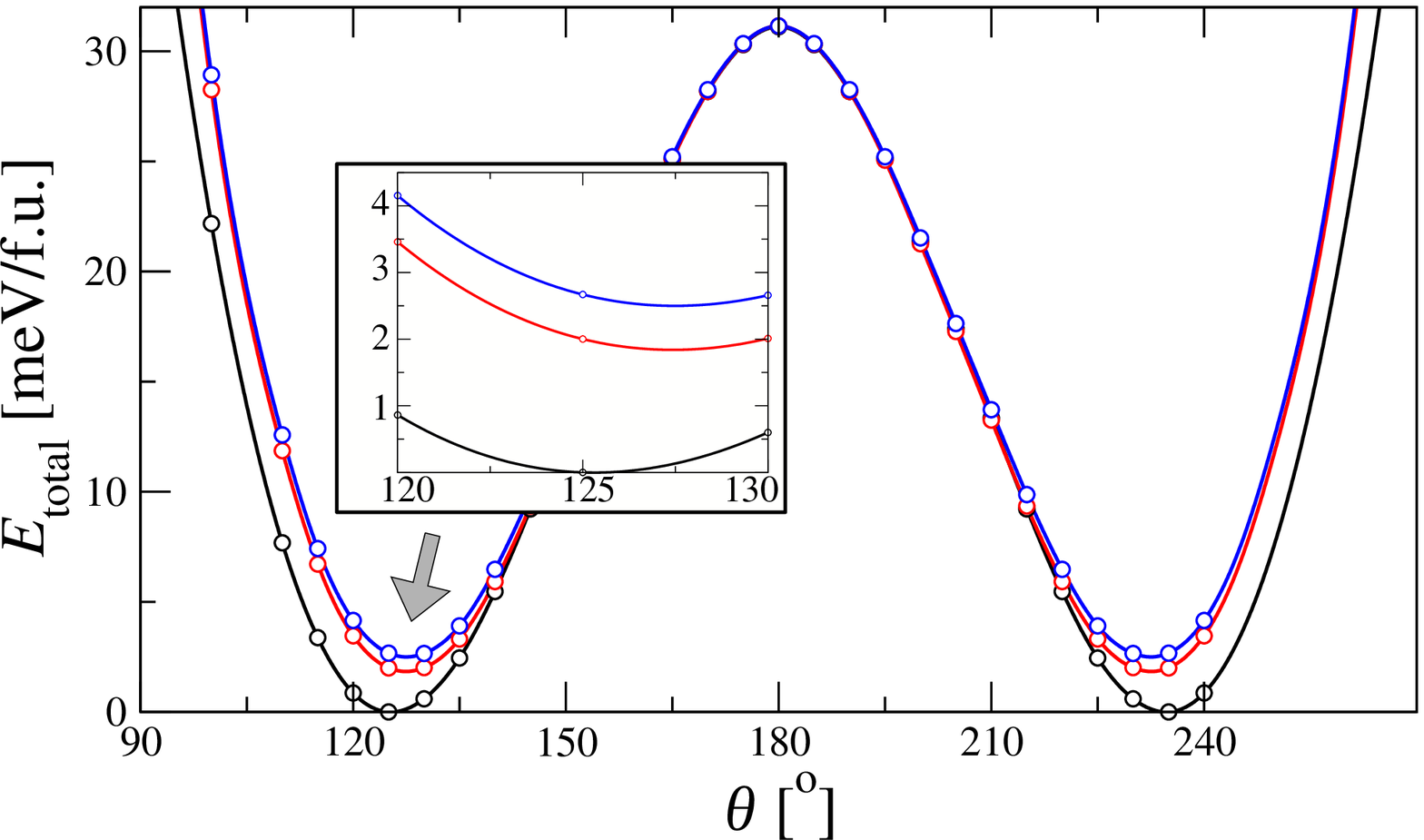}
  \caption{Total energy per Mn$_2$RhSn formula unit computed as a function of
    the orientation of the Mn$_{\rm II}$ magnetic moment characterized by
    angle~$\theta$. We compare three magnetic
    orientations: black - total magnetization is along the $c$-axis; blue and
    red - total magnetization within $ab$-plane, but with Mn$_{\rm
      II}$-moments staggering within $ab$ and within $ac$-planes,
    respectively. Inset shows the detailed energy trends near to the
    canting minimum at about 125$^\circ$.
\label{fig:etotal_theta_mca}}
\end{figure}

In order to analyze the possible long-range magnetic orders, and in
particular, the possibility of skyrmions in Mn$_2$RhSn,
we  computed  the absolute  values  of the  Dzyaloshinskii-Moriya (DM)  vectors  for
Mn$_2$RhSn, by following  the scheme intoduced in Ref.~\cite{EM09}.
In addition we computed the magnetocrystalline anisotropy as the energy
difference between orientations of the total magnetization along the
$c$-axis and within $ab$-plane. Due to canting, for the second case
 we distinguish two orientations - first, when  Mn$_{\rm II}$ moments stagger
 within $ab$-plane and second - within $ac$-plane (see  Fig.~\ref{fig:etotal_theta_mca}).
As it follows from the inset, the energy minima for the
$ab$-orientations are shifted by few degrees.
Their absolute values are about $2.5$  (staggering within $ab$-plane)
and $1.8$ meV/f.u. (staggering within $ac$-plane).

\section{Continuum model of magnetic order in \boldmath ${\rm Mn_2RhSn}$}

The phenomenological continuum theory for the magnetism
of Mn$_2$RhSn can be written in terms of the four sublattices (${l=1,2,3,4}$)
consisting of the two sublattices Mn$_{\,\rm I}$ on Wyckoff site $2a$
and Mn$_{\rm II}$ on site $2c$.
We use standard methods to derive a quantitative model
in the shape of this phenomenological theory by
a systematic coarse graining of a microscopic model,
where direct and antisymmetric DM  exchange
are calculated with DFT methods (discussed in previous section).
Eventually, by adding magnetic anisotropies, also from DFT calculations
and Zeeman energy, a micromagnetic low-temperature continuum
model can be constructed. For the thermal phase diagram, empirical input
is needed to write a Landau-Ginzburg functional for coupled magnetic modes.
As the low acentric symmetry of Mn$_2$RhSn allows for the presence of
chiral inhomogeneous DM couplings, the resulting model
has the form of a Dzyaloshinskii-model~\cite{Dz64} marked by the presence of
Lifshitz-type invariants that couple different magnetic order parameters.

The magnetic moments ${\mathbf S}_l$ in each unit cell of the
lattice ${\mathbf R}_{\mathbf n}, {\mathbf n}=(i,j,k)$
of each sublattice are expressed by
a continuous functions ${\mathbf m}_l({\mathbf r})$
with the property
\begin{equation}
\label{mcontinuum}
{\mathbf m}_l({\mathbf R}_{\mathbf n}+{\mathbf b}_l) = {\mathbf S}_l({\mathbf R}_{\mathbf n})\,,
\end{equation}\
where ${\mathbf b}_l$ are the base vectors of the sublattice sites.

The magnetic free energy is expressed by a standard gradient expansion up to
square terms,
\begin{eqnarray}
\begin{aligned}
w  = &~~\sum_{l,m} \sum_{\alpha\beta} A_{lm}^{\alpha\beta} \partial_{\alpha} {\mathbf m}_l \cdot \partial_{\beta} {\mathbf m}_m\\
   + &~~\sum_{\gamma} \sum_{l,m} \sum_{\alpha\beta}  D_{lm}^{(\gamma)\alpha\beta} m_l^{\alpha}\bar{\partial}_{\gamma}\,m_m^{\beta} \\
   + &~~w_1(\{{\mathbf m}_l\}).
\end{aligned}
\label{w}
\end{eqnarray}
The first two lines describe the inhomogeneous exchange by a set of (anisotropic)
constants $A_{lm}^{\alpha\beta}$, where $\alpha\beta$ are labels of Cartesian coordinates,
and the second line gives the inhomogeneous DM couplings $D_{lm}^{(\gamma)\alpha\beta}$
that arise in low symmetry crystals.
The Lifshitz-type invariants are written
in short form, ${a\bar{\partial}b\equiv a\partial b-b\partial a}$.
Finally $w_1$ collects the terms  which are homogeneous in the set of functions $\{{\mathbf m}_l\}$.
Within our ansatz this contribution can be written as ${w_1 = w_0 + w_a + w_h \dots}$,
where $w_0$ collects contributions deriving from isotropic exchange only,
and $w_a, w_h$ contains anisotropic and Zeeman terms, including the demagnetization energy.

This coarse grained continuum theory for the ground-state can be derived from the
microscopic classical Hamiltonian of the lattice and symmetry constraints.
Using the results of the {\it ab initio} calculations the magnetic energy can be
expressed by a model including direct (isotropic) Heisenberg-like exchange
couplings and the DM-couplings:
\begin{eqnarray}
\begin{aligned}
\!\!\!\!\!H  =  -&\frac{1}{2} \sum_{{\mathbf n}\;{\mathbf p}} \sum_{lm} J_{lm}({\mathbf R}_{\mathbf p} - {\mathbf R}_{\mathbf n})
\;{\mathbf S}_l({\mathbf R}_{\mathbf n})\cdot{\mathbf S}_m({\mathbf R}_{\mathbf p}) \\
 +&  \sum_{{\mathbf n}\;{\mathbf p}} \sum_{lm} \mathbf{D}_{lm}({\mathbf
   R}_{\mathbf p}-{\mathbf R}_{\mathbf n})\!\cdot\!({\mathbf
  S}_l({\mathbf R}_{\mathbf n})\times{\mathbf S}_m({\mathbf
  R}_{\mathbf p}))\,.
\end{aligned}
\label{hHDM}
\end{eqnarray}

Expanding the continuous functions for the sublattices into a Taylor series,
\begin{equation}
\label{mtaylor}
{\mathbf m}_l({\mathbf r}_0-{\mathbf r})={\mathbf m}_l({\mathbf r}_0)+\sum_{\nu} \frac{1}{\nu!}\,[({\mathbf r}-{\mathbf r}_0)\,\cdot\,\nabla]^{\nu}\mathbf{m}_l({\mathbf r}_0)\,,
\end{equation}
and using Eq.(\ref{mcontinuum}) in this Heisenberg-DM-Hamiltonian, the continuum theory can
be derived.
For Mn$_2$RhSn with the tetragonal lattice described by space group $I\bar{4}m2$
crystal class $\bar{4}m2$ ($D_{2d}$), the effective model simplifies to
\begin{eqnarray}
\begin{aligned}
w =&  \sum_{lm}  \sum_{x=a,b,c} A_{lm}^{x} \sum_{\alpha} (\partial_x m_l^{\alpha}\partial_x m_m^{\alpha})\\
  +& \sum_{lm} [ D_{lm}^a  m_l^z \bar{\partial}_a m_m^b + D_{lm}^b m_l^z \bar{\partial}_b m_m^a ]\\
  +&  \sum_{lm} J_{lm} {\mathbf m}_l\cdot{\mathbf m}_m\,,
\end{aligned}
\label{wi}
\end{eqnarray}
%
where the surface terms and constants have been omitted.
The gradients $\partial_a,\partial_b,\partial_d$ are written along and
in units of the tetragonal lattice cell.
The coefficients $J_{lm}$~(Tab.~\ref{tab:J_lm}), $A_{lm}^{x}$~(Tab.~\ref{tab:Aabc_lm}), $D_{lm}^x$~~(Tab.~\ref{tab:Dabc_lm}), now describe effective
coarse grained exchange and DM-couplings.
\begin{table}[htb!]
\begin{tabular}{l|rrrr}
  $l\backslash m$ & 1 & 2 & 3 & 4 \\ \hline
  1     & 7.5 & -10.4 & 4.1 & -10.3 \\
  2     & -10.4 & 21.3 & -10.3 & -27.2 \\
  3     & 4.1 & -10.3 & 7.5 & -10.4 \\
  4     & -10.3 & -27.2 & -10.4 & 21.3
\end {tabular}
\caption{ Coefficients of the effective homogeneous exchange $J_{lm}$~[meV/$\mu_{\rm B}^2$]. \label{tab:J_lm}   }
\end{table}
\begin{table}[htb!]
\begin{tabular}{l|rrrr|rrrr|rrrr}
  & \multicolumn{4}{|c|}{$A^a$} & \multicolumn{4}{|c|}{$A^b$} &  \multicolumn{4}{|c}{$A^c$} \\ \hline
  $l\backslash m$ & 1 & 2 & 3 & 4 & 1 & 2 & 3 & 4 & 1 & 2 & 3 & 4 \\ \hline
  1     & 5.0 & 5.1 & -0.4 & -1.3 & 4.9  & -1.4 & -0.4 & 5.2 & 0.8  & 2.5 & 1.0 & 2.6 \\
  2     & 5.1 & 21.2 & -1.3  & -2.4 & -1.4 & 21.2 & 5.2  & -2.5 & 2.5 & 7.9 & 2.6  & -6.8 \\
  3     & -0.4 & -1.3 & 5.0 & 5.1 & -0.4 & 5.2 & 4.9 & -1.4 & 1.0 & 2.6 & 0.8 & 2.5 \\
  4     & -1.3 & -2.4 & 5.1 &  21.2 & 5.2  & -2.5 & -1.4 &  21.2 & 2.6  & -6.8 & 2.5 &  7.9
\end {tabular}
\caption{ Coefficients of the inhomogeneous exchange $A^{a,b,c}_{lm}$~[meV/$(\mu_{\rm B}^2a^2)$]. \label{tab:Aabc_lm}}
\end{table}
\begin{table}[htb!]
\begin{tabular}{l|rrrr|rrrr}
  & \multicolumn{4}{|c|}{$D^a$} & \multicolumn{4}{|c}{$D^b$}  \\ \hline
  $l\backslash m$ & 1 & 2 & 3 & 4 & 1 & 2 & 3 & 4  \\ \hline
  1     & -0.34 & 0.04  & -0.47  & -0.49 & -0.34  & -0.49  & -0.47 & 0.04 \\
  2     & 0.00  & 0.64  & -0.49  & 0.21 & 0.00 & 0.64 & 0.04  & 0.21  \\
  3     & -0.47  & -0.49 & -0.34  & 0.04 & -0.47  & 0.04 & -0.34 & -0.49\\
  4     & -0.49 & 0.21 & 0.04& 0.64 & 0.04  & 0.21 & -0.49 & 0.64
\end {tabular}
\caption{ Coefficients of the inhomogeneous DM-exchange $D^{a,b}_{lm}$~[meV/$(\mu_{\rm B}^2a)$]. \label{tab:Dabc_lm}}
\end{table}

It must be noted that there is no weak ferromagnetism or weak antiferromagnetism
in the tetragonal inverse Heusler structure, i.e., there are no bilinear coupling
terms between components of the  staggered and ferromagnetic vectors derived from
the DM-exchange, because the four sublattices are related by non-primitive translations.
After quantification of the exchange couplings terms,
it is convenient to analyse the 4-sublattice system by using staggered
and ferromagnetic  vectors of two sublattices,
\begin{eqnarray}
\begin{aligned}
\!\!\!\!m_{\textrm{Mn$\rm_I$}}\!\cdot\!{\mathbf L}=\frac{({\mathbf m}_1-{\mathbf m}_3)}{2}&;&\!\!m_{\textrm{Mn$\rm_I$}}\!\cdot\!{\mathbf F}=\frac{({\mathbf m}_1+{\mathbf m}_3)}{2}; \\
\!\!\!\!m_{\textrm{Mn$\rm_{II}$}}\!\cdot\!{\mathbf l}=\frac{({\mathbf m}_2-{\mathbf m}_4)}{2}&;&m_{\textrm{Mn$\rm_{II}$}}\!\cdot\!{\mathbf f}=\frac{({\mathbf m}_2+{\mathbf m}_4)}{2},
\end{aligned}
\label{MnIMnII}
\end{eqnarray}
where the spin vectors   ${\mathbf L}$ and ${\mathbf F}$ are related to
Mn$\rm_{\,I}$, and ${\mathbf l}$ and ${\mathbf f}$ -- to Mn$_{\rm II}$
sublattice.
In the ground-state configurations, these vectors fulfill
\begin{eqnarray}
\begin{aligned}
{\mathbf L}^2+{\mathbf F}^2=1&;& \;\;\;\;\;{\mathbf L}\cdot{\mathbf F}=0\\
{\mathbf l}^2+{\mathbf f}^2=1&;& \;\;\;\;\;{\mathbf l}\cdot{\mathbf
  f}=0\,.
\end{aligned}
\label{lmunity}
\end{eqnarray}
%
%
Dropping irrelevant terms, the homogeneous part of the continuum theory now is expressed as
\begin{eqnarray}
\begin{aligned}
\tilde{w}_0  = &~~J_{F}{\mathbf F}\cdot{\mathbf F}+J_L{\mathbf L}\cdot{\mathbf L} +J_{f}{\mathbf f}\cdot{\mathbf f}+J_l{\mathbf l}\cdot{\mathbf l}\\
+ &~~J_{c}{\mathbf F}\cdot{\mathbf f} +J'{\mathbf F}\cdot{\mathbf l} \\
- &~~2(m_{\textrm{Mn$\rm_{I}$}}\!\cdot\!{\mathbf
  F}+m_{\textrm{Mn$\rm_{II}$}}\!\cdot\!{\mathbf f})\cdot{\mathbf H}\,,
\end{aligned}
\label{w0}
\end{eqnarray}
with coefficients in~[meV]:
\begin{center}
\begin{tabular}{c|c|c|c|c|c|c|c|c|c}
$J_{F}$ & $J_{L}$ & $J_{f}$ & $J_l$ &
  $J_c$ & $J_{Ff}$ & $J_{Fl}$ & $J_{Lf}$ &
  $J'$ & $J_{Ll}$ \\ \hline
-285.4 & -83.1 & 111.3 & -920.0 & 898.2 & 898.2 & 0 & 0 & 4.8 & 4.8
\end {tabular}\,.
\end{center}
Here we use an obvious notation for the effective exchange, internal to the
ordering modes and between the modes, and using the spin moments from
the DFT calculations ${m_{\textrm{Mn$_{\rm I
        (II)}$}}=3.51~(3.08)~\mu_{\rm B}}$ (see Sec.~\ref{sec:computational}).
The field ${\mathbf H}$ in (\ref{w0}) is the internal magnetic field.
It is seen that the exchange couplings have a clear hierarchy, showing that magnetic
ordering is either dominated by the FM order on Mn$_{\,\rm
  I}$ sublattice or by staggered AFM order on Mn$_{\rm II}$. This AFM order is only very weakly coupled via the staggered vectors ${\mathbf L}$,
which however is not the dominating magnetic mode on Mn$_{\,\rm I}$ sublattice.
Hence, Mn$_2$RhSn is close to a tetracritical (or bicritical) point,
where these two magnetic modes would coexist with the paramagnetic state.
The FM mode ${\mathbf f}$ is a secondary
magnetic order for sublattice Mn$_{\rm II}$, which is \textit{antiparallel} to ${\mathbf F}$ via the very strong coupling $J_c$.
The superposition of ${\mathbf F}$, ${\mathbf l}$ and ${\mathbf f}$ determines a canted state
for the magnetic order with the magnetic cell equivalent to the crystallographic unit cell, i.e. a $\Gamma$-point mode.

\textit{Ground state.}
The homogeneous ground state can be found by neglecting in (\ref{w0}) the small coupling $J'$,
and writing a coplanar canted state arbitrarily in the $ac$-plane,
using ${\mathbf F}\equiv{\mathbf F}_0=(0,0,1)$, ${\mathbf l}\equiv{\mathbf l}_0=(\sin[\pi-\vartheta],0,0)$,
and ${\mathbf f}\equiv{\mathbf f}=(0,0,\cos[\pi-\vartheta])$.
Its energy is given by
\begin{equation}
\label{gsw0}
w_0=J_F + J_l + (J_f-J_l)\,\xi^2+J_c\xi;\;\;\;\; \xi=\cos[\pi-\vartheta]\,.
\end{equation}
The solution for the canting angle is
\begin{equation}
\label{gstheta}
\vartheta=\pi-\arccos\frac{J_c}{2(J_f-J_l)}\,.
\end{equation}
Using the parameters above, the canting angle is ${\vartheta_0=64.2^{\circ}}$,
which is in reasonable agreement with the {\it ab initio} calculations finding
${\theta_{1,2}=55^{\circ}}$, considering that the exchange approximation in Eq.~(\ref{w0})
neglects the anisotropy which has an easy-axis character for
the canted sublattice Mn$_{\rm II}$ (see Fig.~\ref{fig:etotal_theta_mca}).
The ground state, thus, is composed of the two modes ${\mathbf F}$ and ${\mathbf l}$,
which are almost decoupled for ${J'\simeq0}$, and by the induced FM mode ${\mathbf f}$
on Mn$_{\rm II}$. However, the conditions (\ref{lmunity}) provide  a non-linear coupling between
these different modes in a proper micromagnetic model.

{\it Inhomogeneous states.}
The micromagnetic model requires now to include the exchange terms
and the inhomogeneous DM-couplings, i.e., the gradient energy is given
by
\begin{equation}
\label{w2}
w_2 = w_J + w_D,
\end{equation}
where we have the squared gradient terms derived from the isotropic exchange in the form
\begin{eqnarray}
\label{wJ}
\begin{aligned}
w_J =&~A_{F}^{\perp} [ \partial_a {\mathbf F}\cdot \partial_a {\mathbf F} + \partial_b {\mathbf F}\cdot \partial_b {\mathbf F}]
                 + A_{F}^{||} \partial_c {\mathbf F}\cdot \partial_c {\mathbf F}\\
   +&~A_{l}^{\perp} [ \partial_a {\mathbf l}\cdot \partial_a {\mathbf l} + \partial_b {\mathbf l}\cdot \partial_b {\mathbf l}]
                 + A_{l}^{||} \partial_c {\mathbf l}\cdot \partial_c {\mathbf l}\\
   +&~A_{f}^{\perp} [ \partial_a {\mathbf f}\cdot \partial_a {\mathbf f} + \partial_b {\mathbf f}\cdot \partial_b {\mathbf f}]
                 + A_{f}^{||} \partial_c {\mathbf f}\cdot \partial_c {\mathbf f}\\
   +&~A_{Ff}^{\perp} [ \partial_a {\mathbf F}\cdot \partial_a {\mathbf f} + \partial_b {\mathbf F}\cdot \partial_b {\mathbf f}]
                 + A_{Ff}^{||} \partial_c {\mathbf F}\cdot \partial_c {\mathbf f}
\end{aligned}
\end{eqnarray}
with coefficients in [meV/$a^2$] units: 
\begin{center}
\begin{tabular}{c|c|c|c|c|c|c|c|c|c}
$A_{F}^{\perp}$ & $A_{F}^{||}$ & $A_{f}^{\perp}$ & $A_{f}^{||}$ &
  $A_{l}^{\perp}$ & $A_{l}^{||}$ & $A_{Ff}^{\perp}$ & $A_{Ff}^{||}$ &
  $A_{Fl}^{\perp}$ & $A_{Fl}^{||}$ \\ \hline
9.25 & 3.61 & 37.50 & 2.23 & 47.30 & 29.40 & 15.11 & 20.33 & $\simeq$\,0
& $\simeq$\,0  
\end {tabular}\,.
\end{center}

The inhomogeneous DM-couplings are
\begin{eqnarray}
\label{wD}
\begin{aligned}
w_D =&~~D_{F} [ F^c \bar{\partial}_b F^a + F^c \bar{\partial}_a F^b]\\
\label{wDl}
    + &~~D_{l} [ l^c \bar{\partial}_b l^a + l^c \bar{\partial}_a l^b]\\
     + &~~D_{f} [ f^c \bar{\partial}_b f^a + l^c \bar{\partial}_a f^b]\\
     +  &~~D_{c} [ F^c \bar{\partial}_b f^a + F^c \bar{\partial}_a
  f^b+f^c \bar{\partial}_b F^a + f^c \bar{\partial}_a F^b]\,,
\end{aligned}
\label{wDlast}
\end{eqnarray}
with coefficients in [meV/$a$] units:
\begin{center}
\begin{tabular}{c|c|c|c}
$D_{F}$ & $D_{l}$ & $D_{f}$ & $D_{c}$  \\ \hline
-9.8 & 5.1 & 7.0 & -9.8 \\
\end {tabular}\,.
\end{center}
An inhomogeneous modification of the  ground-state takes place on
long lengths, owing to the weakness of the DM-exchange compared
to the direct exchange. Considering that the DM-couplings and also applied magnetic fields and
anisotropies are small in comparison to the strong local exchange forces, the basic
canted structure is preserved in each unit cell. But it can be slowly rotated
over length of many unit cells.

\textit{Landau-Ginzburg functional.}
In order to complete the phenomenological theory, we
briefly discuss the form of an appropriate Landau-Ginzburg functional
which could be used to model the thermal phase diagram and the phase transitions.
The two primary order parameters (OPs) are the FiM and the AFM modes,
${\mathbf F}$ and ${\mathbf l}$.
They and the coupling between them have to be considered with respect to
the secondary OP, which is the FM mode ${\mathbf f}$.
The complete Landau-Ginzburg (LG) functional contains Lifshitz invariants $w_D$ from
Eqs.~(\ref{wD}).
Hence, this LG-functional is not
a simple extension of a proper Landau-theory with the set of applicable
squared-gradient terms $w_J$ from (\ref{wJ}), as the magnetic free energy violates the Lifshitz condition.
Consequently, this \textit{Dzyaloshinskii model} should be understood as
a pseudo-microscopic continuum theory.
Still, the LG functional for Mn$_2$RhSn can be written by using standard Landau
expansion for the homogeneous coupling terms, instead of $\tilde{w}_0$:
\begin{eqnarray}
\begin{aligned}
\label{w0LG}
w_0  = &~a_F {\mathbf F}\cdot{\mathbf F}+b_F ( {\mathbf F}\cdot{\mathbf F})^2\\
    + &~a_l{\mathbf l}\cdot{\mathbf l}+b_f({\mathbf l}\cdot{\mathbf l})^2\\
     + &~a_f {\mathbf f}\cdot{\mathbf f}\\
    + &~c_f{\mathbf F}\cdot{\mathbf f} + c' {\mathbf F}\cdot{\mathbf l}\\
 +&~b_{Ff}|{\mathbf F}|^2\,|{\mathbf f}|^2 + b_{Fl}|{\mathbf F}|^2\,|{\mathbf l}|^2 + b_{fl}|{\mathbf f}|^2\,|{\mathbf l}|^2 \\
    +&~b_c({\mathbf F}\cdot{\mathbf f})^2 + b'({\mathbf F}\cdot{\mathbf l})^2\\
 + &~\textrm{h.o.t.} \\
- &~2 ({\mathbf F}+{\mathbf f})\cdot{\mathbf H}\,,
\end{aligned}
\end{eqnarray}
where all magnetizations ${\mathbf F}$, ${\mathbf f}$, and ${\mathbf l}$
are now considered as variable length 3-component vectors.
The complete LG functional for the magnetic free energy then is given by
\begin{equation}
\label{LG}
w_{LG}=w_0 + w_J+w_D +w_a\,,
\end{equation}
where the $w_a$ collects anisotropic contributions, not contained in the first three terms, i.e.,
the magnetocrystalline and exchange anisotropies.

A complete microscopic derivation of the terms in the LG-functional would
require a detailed finite temperature statistical theory beyond the
input from the DFT-calculations for ground-states. And, for the behavior
at lower temperatures, the higher-order-terms (h.o.t.'s) are
required in the thermodynamic potential.
However, a semi-quantitative model could be written in the usual manner
by restricting the temperature dependence of the model
to the coefficients of the square terms in the primary OPs:
\begin{eqnarray}
\label{LandauCoeffs}
\begin{aligned}
a_F(T)=&~\alpha_F\,(T-T_C^{\,0})\\
a_l(T)=&~\alpha_l\,(T-T_N^{\,0})
\end{aligned}
\end{eqnarray}
In Mn$_2$RhSn the bare FiM Curie-temperature ${T_C^{\,0}\sim280~{\rm K}<T_C}$, and the  ${T_N^{\,0}\sim80}$~K, i.e. close to the onset of the AFM mode
on Mn$_{\rm II}$-sublattice.
These bare or ideal transition temperatures should not deviate
strongly from the observed magnetic transition temperatures,
because the corrections due to DM-exchange~\cite{UKR06} and anisotropy
are expected to remain small.
Magnitude of the remaining coefficients $a_f$ and of the quartic terms with
coefficients $b_{\nu}$ could roughly be fixed to the empirical ordered moments.
Here, we only note that the suppression of the FM mode ${\mathbf F}$
on Mn$_{\,\rm I}$-sublattice below the onset of the AFM signals, a bicritical behavior
with a repulsive interaction, i.e. ${b_{Fl}>0}$, while ${b_{fl}}$ may be small.
This suggests that the thermal magnetic phase diagram of Mn$_2$RhSn is close to
a {\it bicritical} behavior, where the FiM and AFM modes rather compete and inhomogenous textures can occur.

{\textit{ Skyrmions in the ferrimagnetic state}}.
The qualitative discussion of the LG-functional is sufficient to understand
the basic features of inhomogeneous state in the intermediate temperature
range ${T_N^{\,0}<T<T_C}$ where only the FiM collinear state exists.
In that case, the FM mode on the Mn$_{\rm II}$-sublattice is antiparallel
to FM mode on the Mn$_{\,\rm I}$-sublattice, ${{\mathbf f}=-{\mathbf F}}$.
Inserting this into $w_J+w_D$ in Eqs.~(\ref{wJ})-(\ref{wDlast}) and adding
Zeeman term an magnetocrystalline uniaxial anisotropy yields a functional describing
inhomogeneous FiM states in the $ab$-plane :
\begin{eqnarray}
\begin{aligned}
\label{wtildefmi}
\tilde{w}_{\rm FiM}=&~\tilde{A}
 [ \partial_a {\mathbf F}\cdot \partial_a {\mathbf F} + \partial_b {\mathbf F}\cdot \partial_b {\mathbf F}]\\
+&~\tilde{D} [ F^c \bar{\partial}_b F^a + F^c \bar{\partial}_a F^b]\\
-&~2\,(m_{\rm Mn_{I}} - m_{\rm Mn_{II}})\,{\mathbf F}\cdot{\mathbf H}\\
-&~K\,(F^c)^2 \,.
\end{aligned}
\label{wtildefmilast}
\end{eqnarray}
This free-energy functional for the FiM state is exactly equivalent to the
FM Dzyaloshinskii-model for chiral magnets from crystal class $\overline{4}2m$
studied earlier~\cite{BY89,BH94}.
In particular, the solutions for isolated and condensed chiral ``vortices''
and the magnetic phase diagram presented in this pioneering work desribe
what is now known as chiral skyrmions in the FiM state of
acentric Mn$_2YZ$ inverse tetragonal Heusler alloys.

The parameters of this model for Mn$_2$RhSn can be calculated from the microscopic
input as ${\tilde{A}=A_F+A_f-2\,A_{Ff}=16.5}$~[meV$/a^2$] and  ${\tilde{D}=D_F+D_f-D_c=7.0}$~[meV$/a$].
The DFT calculations of anisotropy~(see Sec.~\ref{sec:computational}) suggest an effective easy-axis anisotropy
of the order ${K\simeq2}$~[meV/f.u.].
%
We may assume that the magnetic coupling coefficients do not depend strongly on temperature,
i.e. their temperature dependence should essentially scale only with the square of the saturation magnetization.
This means that all the ratios of coupling terms in the free energy Eq.~(\ref{wtildefmi})
are almost constant with temperature. Then, we may use the coefficients
from the microscopic ground-state calculation to estimate
materials parameters of Mn$_2$RhSn at elevated temperatures and
evaluate the sizes or stability of skyrmions and the phase diagram in Mn$_2$RhSn
in the FiM state.
The chiral magnetic twisting lengths is given by
\begin{equation}
\label{Lambda}
\Lambda=4\pi\,\tilde{A}/\tilde{D},
\end{equation}
which means $\Lambda\sim$~29.5~$a\sim$~130~nm for Mn$_2$RhSn.
The strengths of the easy-axis anisotropy determines whether
a modulated spiral ground-state exists and whether
a field-induced skyrmion lattice in an effective field pointing
along the $c$-axis occurs (see Figs.~9 and 10 in Ref.~\onlinecite{BH94}).
The different cases can be distinguished by the parameter
\begin{equation}
\label{kappa}
\kappa=\pi\,\tilde{D}/4\,\sqrt{\tilde{A}\,K}\,.
\end{equation}
For ${\kappa<1}$, the ground state is collinear as the strong anisotropy suppresses the spiral state.
For ${\kappa>1.14}$, there is a field-induced skyrmion lattice in the magnetic equilibrium
phase diagram for fields along $c$.

\begin{figure}
  \centering
  \includegraphics[clip,width=1.0\linewidth]{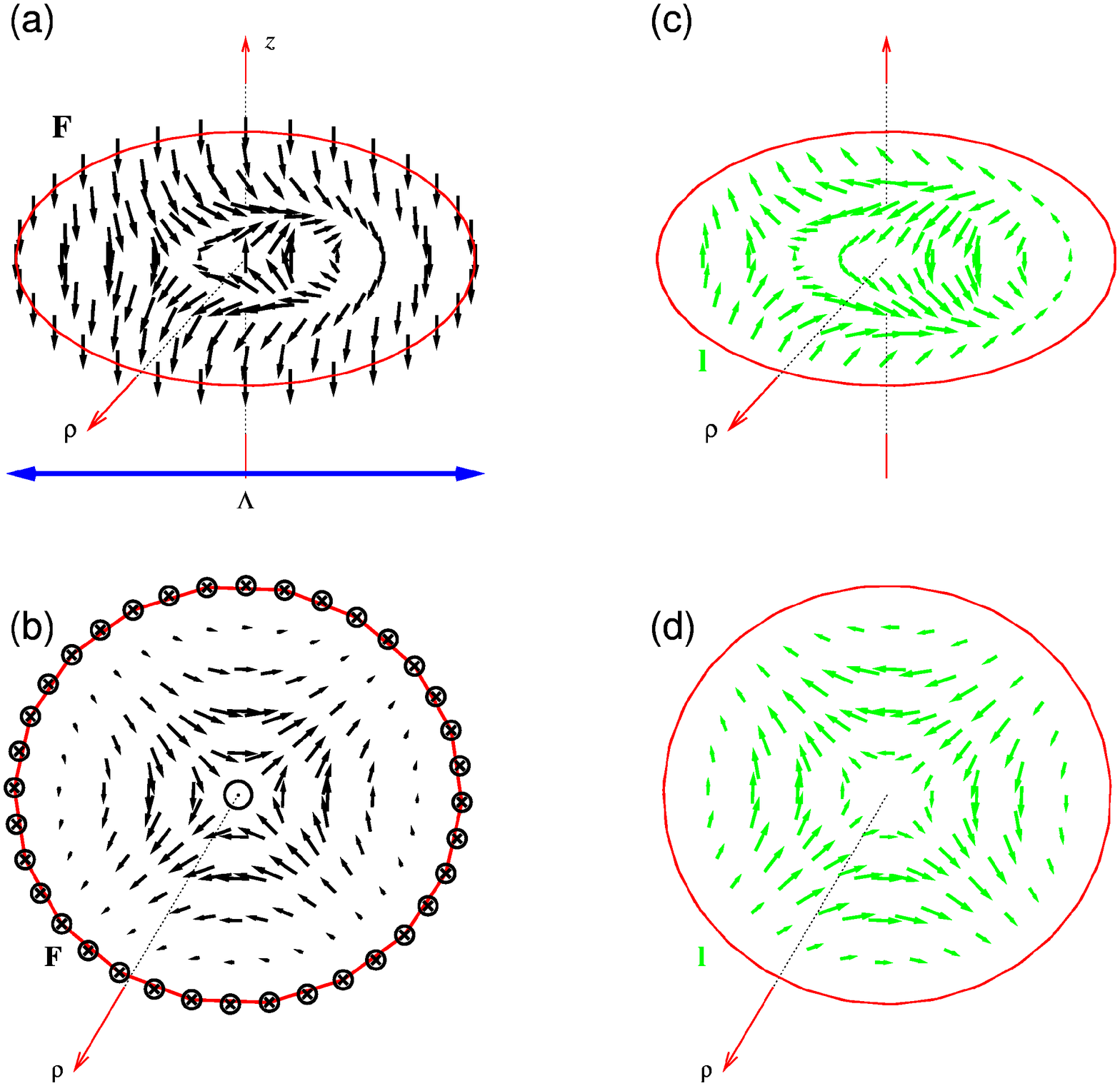}
  \caption{(a)~Shape of the double-twisted skyrmion configuration
in the tetragonal inverse Heusler alloys of $\overline{4}2m$ symmetry.
The FM magnetization ${\mathbf F}$  on sublattice Mn$_{\,\rm I}$
parametrizes the FiM collinear state at higher temperatures
in Mn$_2$RhSn. The corresponding magnetization ${\mathbf f}$ on sublattice Mn$_{\rm II}$
is strictly antiparallel to ${\mathbf F}$.
(b)~Projection of the skyrmion in the $ab$-plane.
(c)~Close to the reorientation transition, the AFM mode ${\mathbf l}$
on Mn$_{\rm II}$-sublattice  sets in: ${|{\mathbf l}|\ll|{\mathbf F}|}$ and also
${|{\mathbf l}|\ll|{\mathbf f}|}$.
${\mathbf l}$ is perpendicular to ${\mathbf F}$ and rotates with it
in the same plane in each radial direction. In the center ${|{\mathbf l}|=0}$.
(d)~Projection of ${\mathbf l}$ onto $ab$-plane.
\label{fig:skyrmions}}
\end{figure}

For the estimated coefficients, we find
${\kappa\simeq0.95}$ as a reasonable value
for Mn$_2$RhSn but close to the critical
$\kappa$.
The spiral magnetic states
in crystals with ${\overline{4}2m}$ crystals
can have demagnetizing fields, as they are
of cycloidal (N{\'e}el)-like character
when propagating along (110)-directions, while
they are of helical (Bloch)-like character for
propagation along (100)-directions.
The demagnetizing field further reduces the
effective $\kappa$ and oblique/skew spirals for propagation
directions in between, as discussed in Ref.~\onlinecite{BH94}.
Hence, the quantitative estimates for the
micromagnetic model $\tilde{w}_{\rm FiM}$ suggest
a collinear FiM state in Mn$_2$RhSn.

The solutions for isolated chiral skyrmions with
a single FM ordering mode have been
presented in Ref.~\onlinecite{BH94} for
the basic model Eqs.~(\ref{wtildefmi}).
The shape of an isolated skyrmion
in Mn$_2$RhSn is sketched in Fig.~\ref{fig:skyrmions}.
Close to the reorientation transition,
where the AFM ordering sets in, the magnitude
of the ${\mathbf l}$-mode is small and can
be modulated (${|\mathbf{l}|\neq}$const). As long this mode is subjugated
to the FiM order it remains perpendicular to the ${\mathbf F}$-mode.
Owing to its softness, it will not only rotate
in a manner, so as to minimize the energy of its Lifshitz-invariants Eq.~(\ref{wDl}),
it also will be modulated with a zero, ${{\mathbf l}=0}$,
in the center in form of a vortex-like defect.
In Mn$_2$RhSn, the coefficients $\tilde{D}$ and
$D_l$ have similar magnitude and the same sign,
so that the screw-sense of the rotation of
${\mathbf F}$ and ${\mathbf l}$ are not in conflict.
Hence, close to $T_N^{\,0}$ the ${\mathbf l}$-mode
follows and co-rotates the FiM mode ${\mathbf F}$
while being modulatd in lengths, Fig.~\ref{fig:skyrmions}~(c, d).
At lower temperatures, the necessity to have
a defect of the ${\mathbf l}$-mode in the skyrmion
center and the associated large defect-energy
most likely will destabilize the skyrmions.
However, there may exist other localized solitonic
textures in this acentric coupled magnetic system,
which may cause inhomogeneous magnetic states to
exist in the acentric Mn$_2YZ$ alloys.

\bibliography{database}